\documentclass[twocolumn]{aastex631}

\usepackage{multirow}
\usepackage{booktabs}
\usepackage{framed}

\shorttitle{Environmental effects in SPT2349$-$56}
\shortauthors{Hughes et al.}

\begin{document}

\title{Evidence for environmental effects in the $z=4.3$ protocluster core SPT2349$-$56}

\author{Chayce Hughes}
\affiliation{University of British Columbia, 6225 Agricultural Road, Vancouver, V6T 1Z1, Canada}

\author{Ryley Hill}
\affiliation{University of British Columbia, 6225 Agricultural Road, Vancouver, V6T 1Z1, Canada}

\author{Scott C. Chapman}
\affiliation{Dalhousie University, 5071 W Saanich Rd, Halifax, B3H 4R2, Canada}
\affiliation{NRC–Herzberg Astronomy and Astrophysics, 5071 W Saanich Rd, Victoria, V9E 2E7, Canada}
\affiliation{University of British Columbia, 6225 Agricultural Road, Vancouver, V6T 1Z1, Canada}

\author{Manuel Aravena}
\affiliation{Instituto de Estudios Astrof\'{\i}sicos, Facultad de Ingenier\'{\i}a y Ciencias, Universidad Diego Portales, Av. Ej{\'e}rcito 441, Santiago, 8320000, Chile}

\author{Melanie Archipley}
\affiliation{Kavli Institute for Cosmological Physics, University of Chicago, 5640 South Ellis Avenue, Chicago, IL 60637, USA}
\affiliation{Department of Astronomy and Astrophysics, University of Chicago, 5640 South Ellis Avenue, Chicago, IL 60637, USA}

\author{Veronica~J. Dike}
\affiliation{Department of Astronomy, University of Illinois, 1002 West Green Street, Urbana, IL 61801, USA}

\author{Anthony Gonzalez}
\affiliation{Department of Astronomy, University of Florida, 211 Bryant Space Science Center, Gainesville, FL 32611-2055, USA}

\author{Thomas R.~ Greve}
\affiliation{Cosmic Dawn Center (DAWN), Technical University of Denmark, DTU Space, Elektrovej 327, 2800 Kgs Lyngby, Denmark}
\affiliation{Department of Physics and Astronomy, University College London, Gower Street, London, WC1E 6BT, UK}

\author{Gayathri Gururajan}
\affiliation{Scuola Internazionale Superiore Studi Avanzati (SISSA), Physics Area, Via Bonomea 265, 34136 Trieste, Italy}
\affiliation{IFPU-Institute for Fundamental Physics of the Universe, Via Beirut 2, 34014 Trieste, Italy}

\author{Chris Hayward}
\affiliation{Center for Computational Astrophysics, Flatiron Institute, 162 Fifth Avenue, New York, NY 10010, USA}

\author{Kedar Phadke}
\affiliation{Department of Astronomy, University of Illinois, 1002 West Green Street, Urbana, IL 61801, USA}

\author{Cassie Reuter}
\affiliation{Department of Physics, University of California, 366 Physics North MC 7300, Berkeley, CA, 94720-7300, USA}

\author{Justin Spilker}
\affiliation{Department of Physics and Astronomy and George P. and Cynthia Woods Mitchell Institute for Fundamental
Physics and Astronomy, Texas A\&M University, 4242 TAMU, College Station, TX 77843-4242, USA}

\author{Nikolaus Sulzenauer}
\affiliation{Max-Planck-Institut f{\"u}r Radioastronomie, Auf dem H{\"u}gel 69, Bonn, D-53121, Germany}

\author{Joaquin~D. Vieira}
\affiliation{Department of Astronomy, University of Illinois, 1002 West Green Street, Urbana, IL 61801, USA}
\affiliation{Center for AstroPhysical Surveys, National Center for Supercomputing Applications, 1205 West Clark Street, Urbana, IL 61801, USA}
\affiliation{Department of Physics, University of Illinois, 1110 West Green St., Urbana, IL 61801, USA}

\author{David Vizgan}
\affiliation{Department of Astronomy, University of Illinois, 1002 West Green Street, Urbana, IL 61801, USA}

\author{George Wang}
\affiliation{University of British Columbia, 6225 Agricultural Road, Vancouver, V6T 1Z1, Canada}

\author{Axel Wei{\ss}}
\affiliation{Max-Planck-Institut f{\"u}r Radioastronomie, Auf dem H{\"u}gel 69, Bonn, D-53121, Germany}

\author{Dazhi Zhou}
\affiliation{University of British Columbia, 6225 Agricultural Road, Vancouver, V6T 1Z1, Canada}

\begin{abstract}

We present ALMA observations of the [C{\sc I}] 492 and 806\,GHz fine-structure lines in 25 dusty star-forming galaxies (DSFGs) at $z\,{=}\,4.3$ in the core of the SPT2349$-$56 protocluster. The protocluster galaxies exhibit a median $L^\prime_{[\text{C{\sc i}}](2-1)}/L^\prime_{[\text{C{\sc i}}](1-0)}$ ratio of 0.94 with an interquartile range of 0.81--1.24. These ratios are markedly different to those observed in DSFGs in the field (across a comparable redshift and 850\,$\mu$m flux density range), where the median is 0.55 with an interquartile range of 0.50--0.76, and we show that this difference is driven by an excess of [C{\sc I}](2--1) in the protocluster galaxies for a given 850\,$\mu$m flux density. Assuming local thermal equilibrium, we estimate gas excitation temperatures of $T_{\rm ex}\,{=}\,59.1^{+8.1}_{-6.8}\,$K for our protocluster sample and $T_{\rm ex}\,{=}\,33.9^{+2.4}_{-2.2}\,$K for the field sample. Our main interpretation of this result is that the protocluster galaxies have had their cold gas driven to their cores via close-by interactions within the dense environment, leading to an overall increase in the average gas density and excitation temperature, and an elevated [C{\sc i}](2--1) luminosity-to-far-infrared luminosity ratio.

\end{abstract}

\keywords{Galaxy environments(2029) --- Protoclusters(1297) --- High-redshift galaxies(734)}

\section{Introduction} \label{sec:intro}

In the local Universe, galaxies residing within galaxy clusters are significantly different from their field counterparts \citep[e.g.,][]{ellis1997,andreon2003,muzzin2012}: cluster galaxies are predominantly red ellipticals that stopped forming stars long ago, while field galaxies are predominantly blue spirals with much higher star-formation rates (SFRs). At high redshift ($z\,{\gtrsim}\,1.5$), when most galaxy clusters are not yet virialized (a regime in which we call these objects protoclusters), quiescent galaxies are much less common within overdense environments \citep[e.g.][]{scoville2013}. Indeed, some observations suggest that protoclusters are actually in an elevated state of star-formation compared to the field \citep[e.g.,][]{overzier2016,chiang2017,popescu2023}, likely driven by mergers and interactions \citep[e.g.,][]{lotz2013,hine2016,andrews2024}. However, the impacts that these processes imprint on protocluster galaxies are not well understood.

The difficulty in directly probing environmental effects on protocluster galaxies stems from the fact that at high redshift, the contrast against the field is typically not large, and subtle differences are often impacted by degeneracies and small sample sizes -- these include differences in stellar ages \citep[e.g.,][]{steidel2005,rettura2010,lee-brown2017,webb2020}, metallicities \citep[e.g.,][]{valentino2015,namiki2019,perez-martinez2024}, and resolved distributions of star formation and stellar mass \citep[e.g.,][]{vulcani2018,cramer2024}, but there is no clear consensus on the importance (or existence, in some cases) of these differences.

The protocluster SPT2349$-$56 was discovered as a bright point source in the South Pole Telescope (SPT) 2500\,deg$^2$ millimetre-wavelength survey \citep{everett2020}, and it was later resolved into a collection of ${>}\,30$ dusty star-forming galaxies (DSFGs) at $z\,{=}\,4.3$ \citep{miller2018,hill2020,rotermund2021,apostolovski2024}.
This system therefore presents an opportunity to directly observe how environmental effects arising from cluster formation are imprinted on cluster galaxies. By combining several continuum observations from the Atacama Large Millimeter/submillimeter Array (ALMA), {\it Spitzer\/}, {\it Gemini\/}, and the {\it Hubble Space Telescope\/} ({\it HST\/}), it appears that the protocluster galaxies in SPT2349$-$56 are consistent with the galaxy main sequence (i.e. field galaxies) around redshift 4 \citep{rotermund2021,hill2022}. On the other hand, detections of the CO(4--3) transition (which is approximately proportional to gas mass) provided tentative evidence that the gas depletion timescales (defined as the ratio of gas mass to SFR) in the SPT2349$-$56 protocluster galaxies are shorter than for field galaxies \citep{hill2022}, but more evidence is needed. 

While observations of the total CO spectral line energy distribution (SLED) would provide more accurate constraints on the total gas mass, lower $J$ lines (especially $J\,{=}\,1$) typically lack the spatial resolution to resolve many of the sources in the crowded SPT2349$-$56 protocluster. However, \citet{zhou2025} recently showed that observations of the CO(4--3) line towards the SPT2349$-$56 protocluster core from the Atacama Compact Array (ACA) recovered a 75\% excess in integrated line strength compared to ALMA observations of the individual galaxies. The low angular resolution of the ACA observations effectively blended all of the protocluster galaxies into a single unresolved sources, meaning that a significant amount of gas (traced by CO(4--3)) must exist as extended emission outside of the individual protocluster galaxies, effectively resolved out by the high angular resolution of ALMA; similar conclusions have been discussed with regard to other protoclusters such as the Distant Red Core (DRC) at $z\,{=}\,4.0$ \citep{oteo2018,ivison2020}.

In this paper we use the neutral carbon atom ([C{\sc i}]) to further probe environmental effects on the properties of the SPT2349$-$56 galaxies. The two neutral atomic carbon fine structure lines ([C{\sc i}](1--0), $\nu_{\rm rest}\,{=}\,$492.2\,GHz, and [C{\sc i}](2--1), $\nu_{\rm rest}\,{=}\,$809.3\,GHz) fall within ALMA's Bands 3 and 4 at $z\,{=}\,4.3$, respectively. Owing to the simple three-level quantum structure of atomic carbon, the ratio of line luminosities can be used to probe the excitation temperature of the [C{\sc i}] gas \citep[e.g.,][]{weiss2003,walter2011,bothwell2017,gururajan2023}. This offers a simple approach to probe the internal state of protocluster galaxies and field galaxies. Moreover, [C{\sc i}] (as an alternative to CO) has been proposed as a tracer of molecular gas \citep[e.g.,][]{weiss2003,papadopoulos2004,dunne2021}, allowing for an independent assessment of the gas content in the SPT2349$-$56 sample. Throughout this paper we assume the $\Lambda$CDM cosmology from \citet{planck2018I} (1\,arcsec$\,{=}\,$6.9\,kpc at $z\,{=}\,$4.3).

\section{Data}\label{sec:data}

\subsection{ALMA observations and data reduction}

The [C{\sc i}](1--0) transition redshifts to 92.9\,GHz at $z\,{=}\,$4.3, which can be observed by ALMA in Band 3. This transition was observed by the ALMA program 2017.1.00273.S (PI S.~Chapman), taken as a two-pointing mosaic targeting all of the known core-component and northern-component protocluster galaxies reported in \citet{hill2020}. The [C{\sc i}](2--1) transition redshifts to ALMA's Band 4 at 152.7\,GHz; the core galaxies were observed by ALMA programs 2018.1.00018.S (PI S.~Chapman) and 2021.1.01313.S (PI R.~Canning), while the northern galaxies were only observed by the latter program. The array configuration used in both Band 4 observing programs were similar and the spectral coverages were configured to be nearly identical. The [C{\sc i}](2--1) line is partially blended (depending on the linewidth) with the CO(7--6) line ($\nu_{\rm rest}\,{=}\,$806.7\,GHz, redshifted to 152.2\,GHz at $z\,{=}\,$4.3), which we also observe for all of our targets.

The data were calibrated with the standard observatory-provided {\tt ScriptForPI} scripts using the versions of {\tt CASA} \citep{mcmullin2007} appropriate to the ALMA cycle. For the Band 4 observations of the core galaxies, we combined the fully-calibrated visibilities from both observing programs into a single measurement set. Data cubes were made using the {\tt CASA} function {\tt tclean} with Briggs weighting and a robust parameter of 0.5, auto-masking, and cleaning down to a threshold of 0.1\,mJy (roughly 2$\sigma$ per channel for all of the cubes). The average synthesized beamsize for the Band 3 data cube sideband containing the line emission of interest was 1.4\,arcsec$\,{\times}\,$1.0\,arcsec and 0.34\,arcsec$\,{\times}\,$0.29\,arcsec for the Band 4 data cube.

\subsection{Line measurements}

Essentially all of the sources are unresolved in the Band 3 data, so a peak-pixel spectrum approach was used to extract spectra from the data cube. Using the redshifts and line widths taken from the best-fit [C{\sc ii}] line profiles from \citet{hill2020}, we averaged over the channels within $2\sigma$ of the expected position of the [C{\sc i}](1--0) line and searched for the brightest pixels near the galaxy positions given in \citet{hill2020}. We then extracted the spectrum at the brightest pixel position. In the Band 4 data the sources are typically resolved, so we designed circular apertures for each source. The center and radius of each aperture was set to enclose the 3$\sigma$ contour of each source in the channel-averaged map, again using the positions of known galaxies from \citet{hill2020} as a prior. We also tested using aperture photometry on the Band 3 data, finding the aperture measurements of the [C{\sc i}](1--0) line did not change the signal compared to the peak pixel measurements while significantly increasing the uncertainties. 

\begin{figure}
    \includegraphics[width = \linewidth]{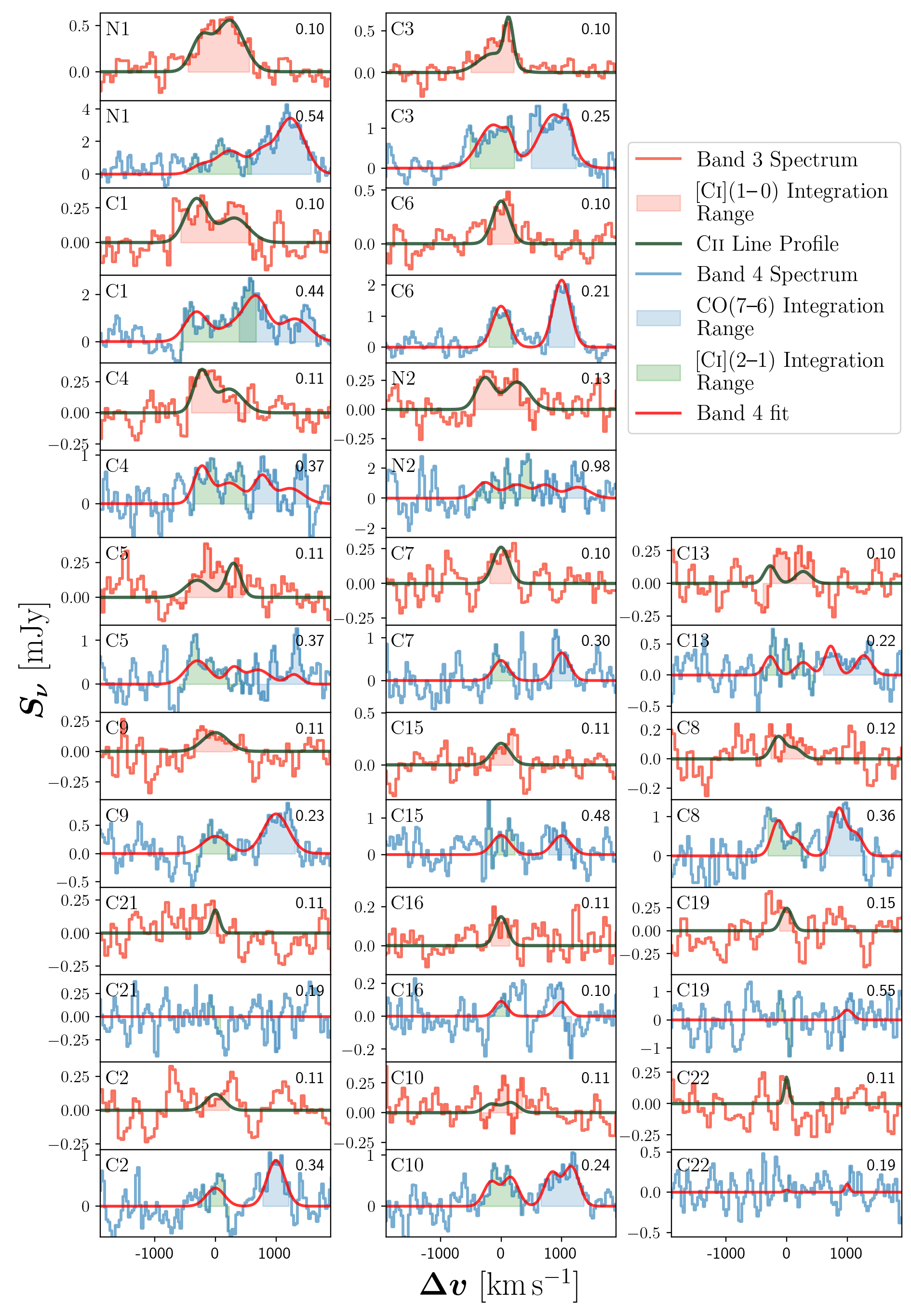}
    \caption{Continuum-subtracted Band 3 and 4 spectra for all sources with a $>2\sigma$ peak pixel [C{\sc i}](1--0) or aperture [C{\sc i}](2--1) measurement. The average per-channel noise in mJy over the displayed channels is listed in the top-right of each panel, and the name of each galaxy in \citet{hill2020} is listed in the top-left. All spectra are unbinned. The fits to the Band 3 and 4 spectra are shown in green and red, respectively. The scaled [C{\sc ii}] line profiles are overlaid on top of the Band 3 spectra. The shaded regions represent the integration ranges, although we note that we subtract the blended Band 4 lines before performing the integration. The sources are ordered by [C{\sc i}](1--0) signal-to-noise ratio.}
    \label{fig:spectra}
\end{figure}

The per-channel noise in the peak pixel spectra was estimated by first masking all the sources with circular masks of the same shape and position as the Band 4 apertures. At each channel, we used sigma clipping with a 3.5$\sigma$ limit to estimate the standard deviation of the background noise pixels. To estimate the per-channel noise in the aperture spectra, we first masked all the sources with circular masks of the same shape and position as the apertures. For each source we then drew 1300 apertures at random positions and then calculated the standard deviation of the 1300 random aperture flux densities. The peak pixel Band 3 and aperture Band 4 spectra of a representative subset of the target sources are shown in Fig.~\ref{fig:spectra}, with the continuum subtracted after fitting the line profiles (as described in detail below).

We deblended the CO(7--6) and [C{\sc i}](2--1) lines in Band 4 using the method described in \citet{chapman2023}. Briefly, the best-fit [C{\sc ii}] line profiles from \citet{hill2020} (where the fit was either a single or double Gaussian) were used as templates, redshifted to the expected frequencies of the CO(7--6) and [C{\sc i}](2--1) lines. For most sources, we fit the [C{\sc ii}] template to the data using three free parameters, namely the CO(7--6) amplitude, the [C{\sc i}](2--1) amplitude, and the continuum level. Sources C3, C5, and N1 required the relative height of the two Gaussians composing their line profiles to be another free parameter. When measuring the line strength of the [C{\sc i}](2--1) line we then subtract the fit to the CO(7--6) line, and vice versa. For completeness we followed the same approach to fit profiles to the [C{\sc i}](1--0) lines, with only two free parameters describing the line height and continuum level. For the brighter sources, we confirmed that the [C{\sc ii}] profiles agreed with the measured CO and C{\sc i} line profiles (see Fig.~\ref{fig:spectra}).

The line strengths were measured by integrating the channels over the region defined by the [C{\sc ii}] line profile: If the [C{\sc ii}] line profile consists of one Gaussian, then the integration region used was ${\pm}\,1.6\sigma$ about the mean. If the [C{\sc ii}] line profile consists of two Gaussians, then the integration region is between $\mu_{\rm left}\,{-}\,1.6\sigma_{\rm left}$ and $\mu_{\rm right}\,{+}\,1.6\sigma_{\rm right}$, where $\mu$ and $\sigma$ are the means and standard deviations of the `left' and `right' Gaussians, respectively. We then scaled the line strengths and errors by the ratio of the total area under the Gaussian to the expected flux inside the integration range, to account for faint flux density missing from the wings of the Gaussian line profiles. For a single Gaussian line profile, this ratio is 1.12. For a double Gaussian line profile, this ratio was calculated for each source (values range between 1.06 and 1.12). We also computed the spatial line emission cutouts for each line, by averaging the data cubes over the line emission channels described above after subtracting the continuum using the {\tt CASA} task {\tt imcontsub} using the remaining channels. We applied the same procedure to the C{\sc ii} data cubes, and in Appendix \ref{AP:line_cutouts} we compare the results; we find that the expected [C{\sc i}](1--0) and [C{\sc i}](2--1) signals in our spectra spatially coincide with the [C{\sc ii}] emission.

Lastly, the integrated line strengths were converted to luminosities using the redshifts taken from the [C{\sc ii}] line centers in \citet{hill2020}. In Table \ref{tab:lines} we provide our results for the [C{\sc i}](1--0), [C{\sc i}](2--1), and CO(7--6) line strengths, as well as 2\,mm continuum flux densities from Band 4 (Band 3 continuum flux densities are already published in \citealt{hill2020}). Where the signal is detected at ${>}\,2\sigma$ we provide the value and the uncertainty, otherwise we provide 3$\sigma$ upper limits.

\text{}\\
\section{Results}\label{sec:results}

\begin{table*}
\centering
\caption{Summary of the observed line strength measurements and the observed flux density at $2000 \, \mu$m, $S_{2000}$. The uncertainties are $1\sigma$. Measurements below $2\sigma$ are shown as $3\sigma$ upper bounds. 
}
\label{tab:lines}
\begin{tabular}{c c c c c c c c}
\toprule
 ID & $\mathrm{L_{[\textsc{ci}](1-0)}}$ & $\mathrm{L_{[\textsc{ci}](2-1)}}$ & $\mathrm{L_{CO(7-6)}}$ & $\mathrm{F_{[\textsc{ci}](1-0)}}$ & $\mathrm{F_{[\textsc{ci}](2-1)}}$ & $\mathrm{F_{CO(7-6)}}$ & $S_{2000}$ \\
 & ($\mathrm{10^7 \, L_\odot}$) & ($\mathrm{10^7 \, L_\odot}$) & ($\mathrm{10^7 \, L_\odot}$) & (Jy\,km\,s$^{-1}$) & (Jy\,km\,s$^{-1}$) & (Jy\,km\,s$^{-1}$) & (mJy) \\
\\
\hline
\multicolumn{8}{c}{Core}\\
\hline
C1 & 3.86 $\pm$ 0.41 & 27.7 $\pm$ 2.4 & 34.7 $\pm$ 2.3 & 0.250 $\pm$ 0.026 & 1.089 $\pm$ 0.094 & 1.371 $\pm$ 0.089 & 1.091 $\pm$ 0.034 \\
C2 & 0.63 $\pm$ 0.28 & 3.4 $\pm$ 1.2 & 7.8 $\pm$ 1.0 & 0.041 $\pm$ 0.018 & 0.132 $\pm$ 0.047 & 0.307 $\pm$ 0.041 & 0.302 $\pm$ 0.024 \\
C3 & 4.01 $\pm$ 0.33 & 17.7 $\pm$ 1.1 & 23.7 $\pm$ 1.0 & 0.259 $\pm$ 0.021 & 0.698 $\pm$ 0.045 & 0.937 $\pm$ 0.040 & 0.899 $\pm$ 0.019 \\
C4 & 3.01 $\pm$ 0.39 & 12.0 $\pm$ 1.8 & 8.9 $\pm$ 1.7 & 0.195 $\pm$ 0.025 & 0.471 $\pm$ 0.071 & 0.351 $\pm$ 0.066 & 0.840 $\pm$ 0.029 \\
C5 & 2.56 $\pm$ 0.39 & 9.0 $\pm$ 1.9 & 7.0 $\pm$ 1.7 & 0.165 $\pm$ 0.025 & 0.355 $\pm$ 0.073 & 0.276 $\pm$ 0.067 & 0.626 $\pm$ 0.029 \\
C6 & 2.12 $\pm$ 0.25 & 11.56 $\pm$ 0.63 & 19.49 $\pm$ 0.67 & 0.137 $\pm$ 0.016 & 0.455 $\pm$ 0.025 & 0.769 $\pm$ 0.026 & 0.624 $\pm$ 0.016 \\
C7 & 1.18 $\pm$ 0.23 & 4.00 $\pm$ 0.89 & 5.67 $\pm$ 0.86 & 0.076 $\pm$ 0.015 & 0.157 $\pm$ 0.035 & 0.224 $\pm$ 0.034 & $<$0.086 \\
C8 & 1.03 $\pm$ 0.30 & 9.8 $\pm$ 1.3 & 12.0 $\pm$ 1.3 & 0.067 $\pm$ 0.020 & 0.387 $\pm$ 0.053 & 0.472 $\pm$ 0.051 & 0.560 $\pm$ 0.027 \\
C9 & 1.31 $\pm$ 0.31 & 3.64 $\pm$ 0.89 & 9.42 $\pm$ 0.87 & 0.085 $\pm$ 0.020 & 0.143 $\pm$ 0.035 & 0.372 $\pm$ 0.034 & 0.097 $\pm$ 0.018 \\
C10 & 0.68 $\pm$ 0.32 & 7.55 $\pm$ 0.89 & 10.90 $\pm$ 0.87 & 0.044 $\pm$ 0.021 & 0.297 $\pm$ 0.035 & 0.430 $\pm$ 0.034 & 0.135 $\pm$ 0.017 \\
C11 & $<$0.801 & $<$3.135 & 3.76 $\pm$ 0.71 & $<$0.052 & $<$0.123 & 0.148 $\pm$ 0.028 & 0.054 $\pm$ 0.016 \\
C12 & $<$0.878 & $<$0.417 & $<$0.513 & $<$0.057 & $<$0.016 & $<$0.020 & $<$0.027 \\
C13 & 0.42 $\pm$ 0.15 & 3.91 $\pm$ 0.91 & 6.06 $\pm$ 0.90 & 0.027 $\pm$ 0.010 & 0.154 $\pm$ 0.036 & 0.239 $\pm$ 0.036 & 0.381 $\pm$ 0.016 \\
C14 & $<$0.879 & $<$1.695 & $<$2.860 & $<$0.057 & $<$0.067 & $<$0.113 & 0.043 $\pm$ 0.016 \\
C15 & 1.10 $\pm$ 0.26 & 6.1 $\pm$ 1.7 & 4.4 $\pm$ 1.5 & 0.071 $\pm$ 0.017 & 0.239 $\pm$ 0.068 & 0.173 $\pm$ 0.059 & 0.158 $\pm$ 0.035 \\
C16 & 0.62 $\pm$ 0.21 & $<$1.245 & $<$1.141 & 0.040 $\pm$ 0.014 & $<$0.049 & $<$0.045 & $<$0.027 \\
C17 & $<$0.564 & $<$1.616 & $<$1.421 & $<$0.036 & $<$0.064 & $<$0.056 & $<$0.041 \\
C18 & $<$0.852 & $<$3.549 & $<$2.091 & $<$0.055 & $<$0.140 & $<$0.083 & $<$0.069 \\
C19 & 0.79 $\pm$ 0.28 & $<$4.522 & $<$4.995 & 0.051 $\pm$ 0.018 & $<$0.178 & $<$0.197 & 0.098 $\pm$ 0.037 \\
C20 & $<$0.734 & $<$1.360 & $<$1.503 & $<$0.048 & $<$0.053 & $<$0.059 & 0.098 $\pm$ 0.013 \\
C21 & 0.43 $\pm$ 0.14 & $<$1.145 & $<$1.140 & 0.028 $\pm$ 0.009 & $<$0.045 & $<$0.045 & 0.046 $\pm$ 0.013 \\
C22 & 0.138 $\pm$ 0.065 & $<$0.806 & $<$1.052 & 0.009 $\pm$ 0.004 & $<$0.032 & $<$0.042 & $<$0.055 \\
C23 & $<$0.713 & $<$1.039 & $<$1.039 & $<$0.046 & $<$0.041 & $<$0.041 & $<$0.039 \\
\hline
\multicolumn{8}{c}{North} \\
\hline
N1 & 6.68 $\pm$ 0.37 & 25.4 $\pm$ 2.6 & 59.7 $\pm$ 2.6 & 0.432 $\pm$ 0.024 & 0.998 $\pm$ 0.101 & 2.356 $\pm$ 0.101 & 1.616 $\pm$ 0.043 \\
N2 & 3.12 $\pm$ 0.45 & 23.0 $\pm$ 4.6 & 13.9 $\pm$ 4.6 & 0.202 $\pm$ 0.029 & 0.905 $\pm$ 0.183 & 0.547 $\pm$ 0.182 & 0.988 $\pm$ 0.078 \\
\bottomrule
\end{tabular}
\end{table*}

To compare our measured line strengths with samples of field DSFGs at similar redshifts as SPT2349$-$56 we use $S_{850}$, the flux density at 850\,$\mu$m, from \citet{hill2020}. $S_{850}$ is model-independent, unlike quantities such as the far-infrared luminosity ($L_{\rm FIR}$), and for $z\,{\gtrsim}\,2$ is roughly independent of redshift. We note that while current dust continuum measurements of the protocluster galaxies in SPT2349$-$56 do not reach the peak of the SED at wavelengths around 500\,$\mu$m, constraints on the average dust temperature of the protocluster population have been estimated to be about 40\,K \citep{hill2020,rotermund2021}, although the uncertainties are still large. Additionally, high-$J$ CO lines (such as $J\,{=}\,7$) have been suggested as tracers of $L_{\rm FIR}$ \citep{bayet2009,liu2015}, however the uncertainties in these scaling relations are comparable to the existing uncertainties in the dust temperatures.

\begin{figure*}
    \centering
    \includegraphics[width = \textwidth]{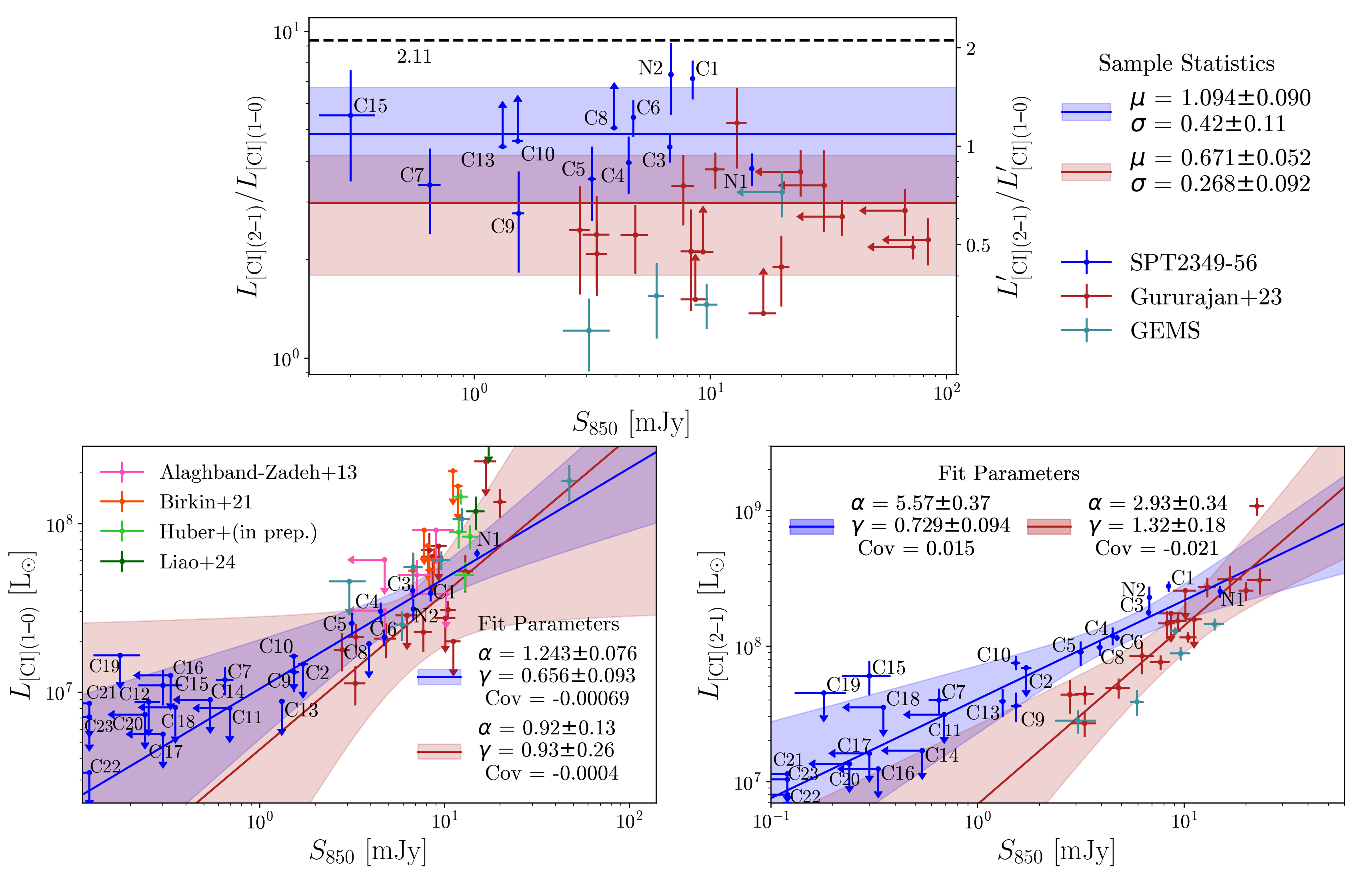}
    \caption{Comparison of [C{\sc i}] and CO line luminosities and line ratios in SPT2349$-$56 and the literature field sample. {\it Top:} The $L_{[\text{C{\sc i}}](2-1)}\,{/}\,L_{[\text{C{\sc i}}](1-0)}$ ratio as a function of $S_{850}$ for protocluster galaxies (blue) and field galaxies (red), with $L^\prime$ units shown on the right axis. The horizontal dashed line at $L^\prime_{[\text{C{\sc i}}](2-1)}\,{/}\,L^\prime_{[\text{C{\sc i}}](1-0)} = 2.11$ indicates where the line ratio becomes unphysical (see Eq.~\ref{eq:Tex}). Sources with ${>}\,3.5\sigma$ measurements in both lines are shown as detections, otherwise we show upper/lower limits. The mean and scatter of the samples are shown as the solid lines and shaded regions, respectively. {\it Bottom left:} $L_{[\text{C{\sc i}}](1-0)}$ versus $S_{850}$ for protocluster galaxies versus field galaxies. Upper limits are shown for measurements below $3.5\sigma$. Best-fit power law models to both SPT2349$-$56 (blue) and \citet{gururajan2023} (red) samples are shown as the solid lines, with the fit parameters and covariances listed. The shaded regions correspond to 95\% confidence intervals assuming the fit parameters $\alpha$ and $\gamma$ follow a multivariate normal distribution. {Bottom right:} Same as bottom left but for $L_{[\text{C{\sc i}}](2-1)}$.} 
    \label{fig:linestrength}
\end{figure*}

In Fig.~\ref{fig:linestrength} we therefore plot our measured line luminosities as a function of $S_{850}$ in three different ways: the top panel shows the ratio $L_{[\text{C{\sc i}}](2-1)}\,{/}\,L_{[\text{C{\sc i}}](1-0)}$, the ratio of the [C{\sc i}] line luminosities, while the bottom-left and bottom-right panels show $L_{[\text{C{\sc i}}](1-0)}$ and $L_{[\text{C{\sc i}}](2-1)}$, respectively. In the line ratio plot we also show the axis converted to $L^\prime_{[\text{C{\sc i}}](2-1)}\,{/}\,L^\prime_{[\text{C{\sc i}}](1-0)}$ (i.e. in units of K\,km\,s$^{-1}$\,pc$^{2}$), which are simply the original units multiplied by a factor of $\left(\nu_{[\text{C{\sc i}}](1-0)}\,{/}\,\nu_{[\text{C{\sc i}}](2-1)}\right)^3$. In all panels we show these quantities as a function of $S_{850}$. In this plot we limit detections to ${>}\,3.5\sigma$ and otherwise show $3\sigma$ upper/lower limits (depending on the quantity being plotted).

Our primary field comparison sample comes from \citet{gururajan2023}, which targeted [C{\sc i}](1--0) and [C{\sc i}](2--1) in a sample of 30 strongly-lensed sources selected from the SPT sample \citep{reuter2020}. The \citet{gururajan2023} sample includes a single blended measurement of SPT2349$-$56, which we discard. Additionally, \citet{gururajan2023} discard SPT0452$-$52 due to an ambiguous redshift at the time, leaving a sample of 28 DSFGs, 17 of which have good lens models \citep{spilker2016}. CO(7--6) and CO(4--3) line luminosity measurements for this sample are also available (Gururajan, priv. comm.). [C{\sc i}](2--1) and CO(7--6) deblending was done in a comparable way, and we checked that our deblending algorithm (described above) applied to their data resulted in similar measurements as reported in their paper. We also include measurements of the two atomic carbon lines from four strong lenses in {\it Planck\/}'s dusty Gravitationally Enhanced sub-Millimetre Sources \citep[GEMS;][]{nesvadba2019}, as well as CO(7--6) and CO(4--3) line luminosity from \citet{canameras2018}; all of these sources have 
lens models \citep{canameras2015}. For completeness, in our comparison of [C{\sc i}](1--0) line luminosities we include field samples from \citet{birkin2021}, \citet{alaghband-zadeh2013}, Huber et al.~(in prep.), and \citet{Liao2024}, although these samples do not have corresponding [C{\sc i}](2--1) measurements. For the \citet{gururajan2023} and \citet{birkin2021} samples, we use $S_{870}$ instead of $S_{850}$ because $S_{850}$ measurements were unavailable. Although we compare with the other samples above, only the \citet{gururajan2023} and GEMS samples have both [C{\sc i}](1--0) and [C{\sc i}](2--1) measurements. 

\citet{walter2011} also observed the [C{\sc i}](1--0) and [C{\sc i}](2--1) lines in a sample of DSFGs and active galactic nuclei (AGN) at a median redshift of 2.8. They found a mean C{\sc i} line ratio of $0.55\pm0.15$ (in $L^\prime$ units), comparable to the \citet{gururajan2023} sample (see Fig. \ref{fig:linestrength}). However, we do not include this literature sample here to avoid confusing the results with the inclusion of AGN and because the median redshift of the sample is much lower than our sample.

The GEMS sample appears to have a lower $L_{[\text{C{\sc i}}](2-1)}\,{/}\,L_{[\text{C{\sc i}}](1-0)}$ ratio than the \citet{gururajan2023} sample. Therefore, we only include the \citet{gururajan2023} sample in the fits and statistics calculations we describe below to have a consistent comparison across all plots and to avoid exaggerating the difference in the $L_{[\text{C{\sc i}}](2-1)}\,{/}\,L_{[\text{C{\sc i}}](1-0)}$ ratio between the SPT2349$-$56 and field samples. 

Next, we compute the means and standard deviations of the line ratios of the two samples, focusing on $L^\prime$ units which are more widely used in the literature; the results are shown in Fig.~\ref{fig:linestrength}. We include the \citet{gururajan2023} sources without lens models in this calculation, under the assumption that both lines are magnified equally. We draw $10^5$ realizations of $[\text{C}\textsc{i}](1-0)$ and $[\text{C}\textsc{i}](2-1)$ line measurements from Gaussian distributions using the line measurement errors, giving $10^5$ realizations of $L^\prime_{[\text{C{\sc i}}](2-1)}\,{/}\,L^\prime_{[\text{C{\sc i}}](1-0)}$ values for each source. Each realization has a mean and standard deviation for the protocluster and field sample from which we calculate an average and spread. We find a significantly higher mean atomic carbon line luminosity ratio in the SPT2349$-$56 sample ($\mu\,{=}\,1.094\,{\pm}\,0.090$) compared to the field sample ($\mu\,{=}\,0.671\,{\pm}\,0.052$), with standard deviations of $\sigma\,{=}\,0.42\,{\pm}\,0.11$ and $\sigma\,{=}\,0.268\,{\pm}\,0.092$ for the protocluster sample and the field galaxy sample, respectively. Additionally, we find a significantly higher median line luminosity ratio in the SPT2349$-$56 sample (median ${=}\,0.94$, interquartile range ${=}\,0.81$--1.24) compared to the field sample (median ${=}\,0.55$, interquartile range ${=}\,0.50$--0.76). 

For completeness, for the $L^\prime_{{\rm CO}(7-6)}\,{/}\,L^\prime_{{\rm CO}(4-3)}$ ratios we find a mean of $\mu\,{=}\,0.604\,{\pm}\,0.029$ for the protocluster sample and $\mu\,{=}\,0.466\,{\pm}\,0.033$ for the field sample (with $\sigma\,{=}\,0.235\,{\pm}\,0.037$ and $\sigma\,{=}\,0.281\,{\pm}\,0.088$, respectively). The $L^\prime_{{\rm CO}(7-6)}\,{/}\,L^\prime_{{\rm CO}(4-3)}$ medians are 0.55 and 0.39, with interquartile ranges of 0.48--0.68 and 0.33--0.47. These results suggest a potential difference in physical properties such as the gas kinetic temperature and density, however a more detailed analysis modeling the full suite of available CO data will be carried out in a future work (Sulzenauer et al.~in prep.).

Turning to the [C{\sc i}](1--0) and [C{\sc i}](2--1) line luminosity versus $S_{850}$ (Fig.~\ref{fig:linestrength}), we fit power laws of the form $L = \alpha \mathrm{C_y} (S_{850}/\mathrm{S_x})^\gamma$ to the protocluster and field samples separately, where $\mathrm{C_y}$ is fixed to 3$\,{\times}\,10^7\,$L$_{\odot}$ and $\mathrm{S_x}$ to 7\,mJy so that our fit parameters are unitless and of order unity. The fits were done with an orthogonal distance regression using {\tt scipy}'s {\tt odr} module. In Fig.~\ref{fig:linestrength} the fits to the field sample are shown in red and the fits to the SPT2349$-$56 galaxies are shown in blue, each with the 95\% confidence region shaded in. The values and $1\sigma$ uncertainties in the fit parameters are also shown in the figure. We find similar fit parameters to the [C{\sc i}](1--0) line luminosities in both populations, while the amplitude of the [C{\sc i}](2--1) fit is significantly larger for the protocluster sample versus the field sample. 

To compare the power law fits, we calculated the average [C{\sc i}] line luminosity ratios (defined as $r\,{=}\,L_\mathrm{C}\,{/}\,L_\mathrm{F}$, where `C' denotes cluster and `F' denotes field) predicted by the fits over the interval between 2 and 20 mJy, where most of the data is situated. As the data is uniformly distributed logarithmically with respect to $S_\mathrm{850}$, we calculated $\bar{r}$ with respect to $\log \left(\frac{S_\mathrm{850}}{\mathrm{S_x}}\right)$ as:
\begin{equation}
    \bar{r} = \frac{ \alpha_\mathrm{C}}{\alpha_\mathrm{F} \log\left( \frac{20}{2} \right) } \int_{\log \left(\frac{2 \, \text{mJy}}{\mathrm{S_x}} \right)}^{\log \left (\frac{20 \, \text{mJy}}{\mathrm{S_x}} \right)} 10^{(\gamma_\mathrm{C} - \gamma_\mathrm{F}) u} du
\end{equation}
\noindent
where $u = \log \left(\frac{S_\mathrm{850}}{\mathrm{S_x}}\right)$. We find $\bar{r}_{10}\,{=}\,1.41\,{\pm}\,0.26$ for [C{\sc i}](1--0) and $\bar{r}_{21}\,{=}\,2.17\,{\pm}\,$0.38 for [C{\sc i}](2--1), suggesting an excess of [C{\sc i}](2--1)$ \, / \, S_\mathrm{850}$ in SPT2349$-$56 compared to the field. Additionally, we find $\frac{\bar{r}_{21}}{\bar{r}_{10}}\,{=}\,1.54\,{\pm}\,0.39$ which is consistent with the ratio of the mean $L_{[\text{C{\sc i}}](2-1)}\,{/}\,L_{[\text{C{\sc i}}](1-0)}$ values between the field and SPT2349$-$56, $1.63\,{\pm}\,0.18$.  

Being a simple three-level quantum system and assuming local thermal equilibrium, the [C{\sc i}](1--0) and [C{\sc i}](2--1) line luminosities can be used to calculate the gas excitation temperature, $T_{\rm ex}$, as \citep{schneider2003}:
\begin{equation} \label{eq:Tex}
    T_\mathrm{ex} =  \frac{38.8\,\mathrm{K}}{\ln \left( \frac{2.11}{R_\mathrm{CI}}\right)},
\end{equation}
\noindent
with $R_\mathrm{CI}$ in the above equation given by 
\begin{equation} \label{eq:R_CI}
    R_\mathrm{CI} =  \frac{L^\prime_{[\text{C{\sc i}}](2-1)}}{L^\prime_{[\text{C{\sc i}}](1-0)}}.
\end{equation}
\noindent
We note that these equations assume that both C{\sc i} lines are optically thin and that the emission arises from photon-dominated regions (PDRs). We find optical depths of ${<}5\,{\times}\,10^{-3}$ for both C{\sc i} lines across our sample, sufficiently thin to apply Eq.~\ref{eq:Tex}. In principle the C{\sc i} gas might not be in thermal equilibrium, and so knowledge of the gas density would be required to estimate the gas temperature more precisely \citep{papadopoulos2022}; however, we regardless expect a positive correlation between the line ratios and gas temperatures for our samples of protocluster and field DSFGs.

Using the mean $R_\mathrm{CI}$ for SPT2349$-$56, we find $T_{\rm ex}\,{=}\,59.1_{-6.8}^{+8.1}$\,K. On the other hand, using the mean $R_\mathrm{CI}$ from the \citet{gururajan2023} sample, we find $T_{\rm ex}\,{=}\,33.9_{-2.2}^{+2.4}$\,K; here the errorbars are 68\% confidence intervals, propagated from the (Gaussian) uncertainties in the mean line luminosity ratios. It is worth pointing out that one of the galaxies in SPT2349$-$56 (C6) has been detected in several radio bands, making it a likely AGN \citep[][]{chapman2023}. Neutral carbon line emission from AGN tend to come from X-ray-dominated regions (XDRs) as opposed to PDRs, the latter of which is assumed in deriving Eq.~\ref{eq:Tex}. However, the line luminosity ratio for C6 is by no means an outlier in our sample, and removing it from our sample has no effect on our sample statistics, so we proceed with the results above. Neutral carbon line luminosity ratios $L^\prime_{[\text{C{\sc i}}](2-1)}\,{/}\,L^\prime_{[\text{C{\sc i}}](1-0)}\,{\gtrsim}\,0.8$ have been suggested to be an indication of AGN due to X-ray heating \citep[e.g.][]{meijerink2007}, although further observational evidence would be required to confirm additional AGN in the SPT2349$-$56 protocluster.

Lastly, it is worth pointing out that additional trends in the $L^\prime_{[\text{C{\sc i}}](2-1)}\,{/}\,L^\prime_{[\text{C{\sc i}}](1-0)}$ ratio might be present. Physical protocluster galaxy size estimates from 870\,$\mu$m dust continuum and [C{\sc ii}] line emisison observations are available from \citet{hill2020} and \citet{hill2022}, respectively, however no correlations were seen. We also looked for correlations between the line strength ratio and distance from the protocluster center, but again we did not find any trends.

\section{Discussion and conclusions}\label{sec:discussion}

Our main finding is an excess in the [C{\sc i}](2--1) line luminosities in the protocluster galaxies found in SPT2349$-$56 compared to field DSFGs at similar redshifts (for a given 850\,$\mu$m flux density), while the [C{\sc i}](1--0) line luminosities are comparable. The difference means that protocluster galaxies have a higher gas excitation temperature ($T_{\rm ex}\,{=}\,59.1_{-6.8}^{+8.1}$\,K) compared to field galaxies ($T_{\rm ex}\,{=}\,33.9_{-2.2}^{+2.4}$\,K). 

A key aspect is that our findings are not very sensitive to the actual $L_{\rm FIR}$ values of the galaxies in either sample. If the dust temperatures of the protocluster galaxies in SPT2349$-$56 are higher than the field galaxies in our comparison sample, then we would interpret the higher line ratios as a deficit in [C{\sc i}](1--0) (as opposed to an excess in [C{\sc i}](2--1)), but our results based on the line ratios would not change. In addition, we note that \citet{hill2022} found evidence that the gas depletion
timescales (effectively the ratio of CO(4–-3) line luminosity to $L_{\rm FIR}$) of the protocluster galaxies in SPT2349$-$56 are smaller than for field galaxies, consistent with them having higher radiation intensity per unit gas mass, and therefore higher excitation temperatures. Our result is also robust against potential beam effects as the smaller Band 4 beam would tend to over-resolve flux compared to the larger Band 3 beam leading to a smaller line ratio, and we have tested several line strength measurement techniques. We have also tested for the potential of spectral confusion between [C{\sc i}](2--1) and CO(7--6) by running our algorithm directly on the comparison field sample.

\citep{cortzen2020} also measured the [C{\sc i}](1--0) and [C{\sc i}](2--1) transitions in GN20, a well-studied DSFG ($S_{850}\,{\approx}\,20\,$mJy, \citealt{pope2006}) at $z\,{=}\,4.05$. This galaxy is potentially the central site of a protocluster \citep{daddi2009}, although the number of spectroscopically-confirmed DSFGs is four, much less than the number of DSFGs found in SPT2349$-$56, and the luminosity distribution is heavily weighted towards the extremely luminous GN20. Nonetheless, \citep{cortzen2020} found a high $L^\prime_{[\text{C{\sc i}}](2-1)}/L^\prime_{[\text{C{\sc i}}](1-0)}$ ratio of 0.94$\,{\pm}\,0.18$, corresponding to a [C{\sc i}] gas excitation temperature of 48$_{-9}^{+14}$\,K and comparable to the mean value found in SPT2349$-$56.

Several interpretations are possible. Simulations show that close-by interactions can drive gas towards the centers of galaxies, thereby increasing the observed average density and gas excitation temperature \citep[e.g.,][]{moreno2015,blumenthal2018,moreno2019}. Alternatively, the protocluster galaxies in SPT2349$-$56 could have had their less dense outer regions (traced by $L_{[\text{C{\sc i}}](1-0)}$) stripped during these interactions, leaving an excess abundance of more centrally-concentrated gas (traced by $L_{[\text{C{\sc i}}](2-1)}$).

It is worth noting that \citet{hill2022} compared the gas mass-to-stellar mass ratios (with gas mass scaled from CO(4--3) detections) of the protocluster galaxies in SPT2349$-$56 to field galaxies found around the same redshift, finding no significant difference between the two populations. Thus to explain our observations with the gas stripping scenario, cold gas (with normal $L_{[\text{C{\sc i}}](2-1)}\,{/}\,L_{[\text{C{\sc i}}](1-0)}$ ratios, similar to field galaxies) must be feeding the core of SPT2349$-$56 to provide an initial excess in gas mass, which is subsequently lost to the forming intergalactic medium. While both the gas stripping and the centrally-concentrated gas scenarios are possible, the latter is simpler and therefore our preferred explanation.

\begin{figure}
    \centering
    \includegraphics[width = \columnwidth]{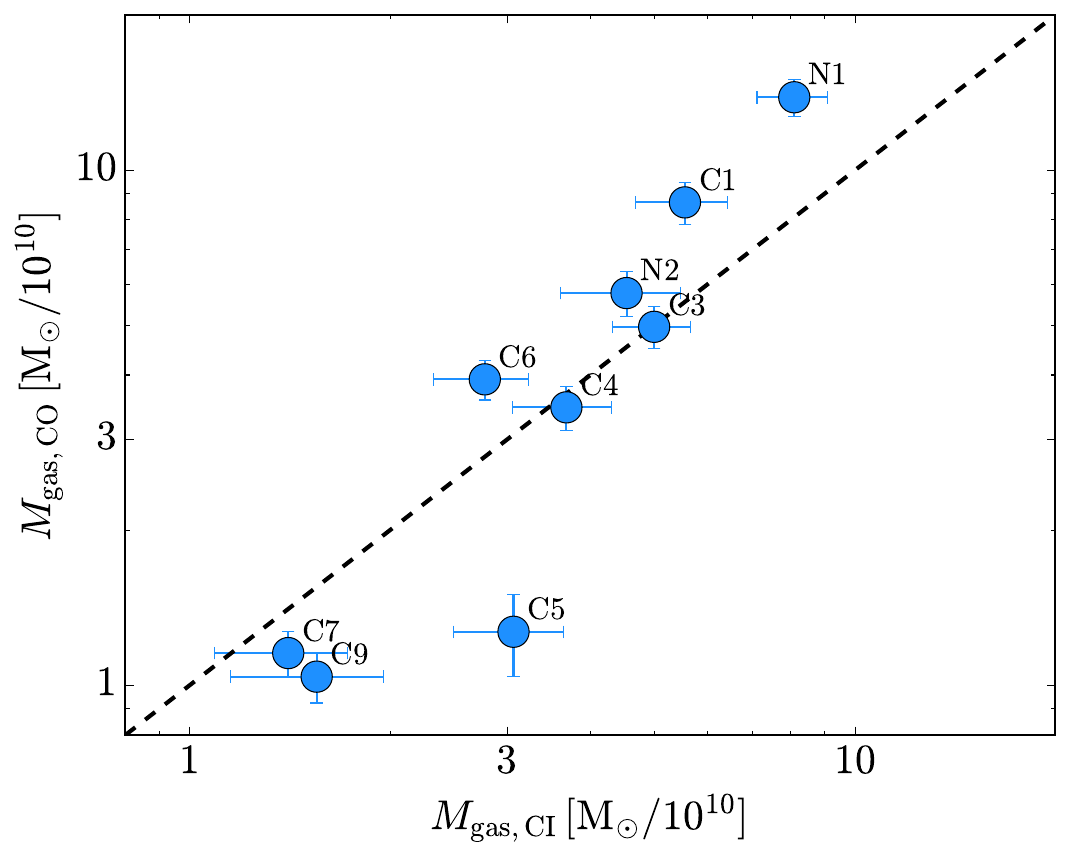}
    \caption{Gas mass estimates from CO(4--3) measurements compared to gas mass estimates from [C{\sc i}]. \citet{hill2020} assumed $r_{4,1}\,{=}\,0.60\pm0.05$ to convert CO(4--3) to CO(1--0) line luminosities (in units of K\,km\,s$^{-1}$\,pc$^2$) and $\alpha_{\rm CO}\,{=}\,1.0\,$M$_{\odot}\,$(K\,km\,s$^{-1}$\,pc$^2$)$^{-1}$. We have used Eq.~\ref{eq:Mgas} with $X_{\rm CI}\,{\times}\,\alpha_{\rm CO}\,{=}\,(6.31\pm0.67)\,{\times}\,10^{-5}$ \citep{gururajan2023}, along with the same $\alpha_{\rm CO}$ value.} 
    \label{fig:Mgas}
\end{figure}

Atomic carbon line luminosities have also been proposed as a gas mass tracer (comparable to CO; e.g., \citealt{papadopoulos2004a,papadopoulos2004,bothwell2017,dunne2021}), and so it is worth investigating how robust previous gas mass estimates are. The molecular gas mass ($M_\mathrm{H_2}$) can be written as \citep{gururajan2023}
\begin{equation}\label{eq:Mgas}
    M_{\rm H_2} = \frac{\mathrm{k} L_\mathrm{[CI](1\text{-}0)}}{Q_{10}\mathrm{X_{CI}}}\,[\mathrm{M_{\odot}}],
\end{equation}
\noindent
where
\begin{equation} \label{eq:Q10}
    \mathrm{\textit{Q}_{10} = \frac{3\exp(-\frac{T_1}{\textit{T}_{ex}})}{1+3\exp(-\frac{T_1}{\textit{T}_{ex}}) + 5\exp(-\frac{T_2}{\textit{T}_{ex}})}},
\end{equation}
\noindent
$\mathrm{k}\,{=}\,3.39\,{\times}\,10^{-2}\,{\rm M}_\odot\,{\rm L}_\odot^{-1}$, $T_1\,{=}\,23.6\,$K, $T_2\,{=}\,62.5\,$K, and $X_{\rm CI}$ is the calibration factor (equal to the neutral [C{\sc i}]/[H$_2$] abundance ratio). \citet{gururajan2023} combined their sample with other literature DSFGs to derive a mean calibration factor of $X_{\rm CI}\,{\times}\,\alpha_{\rm CO}\,{=}\,(6.31\pm0.67)\,{\times}\,10^{-5}$, which we apply to our sample along with $\alpha_{\rm CO}\,{=}\,1.0\,$M$_{\odot}\,$(K\,km\,s$^{-1}$\,pc$^2$)$^{-1}$ (as was used by \citealt{hill2020}). In Fig.~\ref{fig:Mgas} we compare the gas masses from \citet{hill2020} (derived from CO(4--3), assuming $r_{4,1}\,{=}\,0.60\pm0.05$) to gas masses derived from [C{\sc i}] (using Eq.~\ref{eq:Mgas}). We find good agreement between the overall calibration factors applied to the two mass estimates, with a small systematic error leading to a slight tilt in the $M_{\rm gas,CI}$--$M_{\rm gas,CO}$ relation. However, this tilt only leads to differences between the two gas mass tracers within a factor of 2.

The two main interpretations we have discussed regarding the difference in [C{\sc i}] line luminosity ratios between protocluster galaxies and field galaxies at high redshift, namely nearby interactions either driving cold gas to the cores of the protocluster galaxies or stripping gas from the less dense outskirts, cannot be conclusively distinguished with our current data. 
Further ALMA observations directly measuring the far-infrared continuum (thus constraining the overall dust continuum temperatures), along with bolstering the result with lower $J\,{=}\,1\text{--}0$ and 2--1 CO measurements of the cold gas, will allow a more complete explanation of the results. Measuring additional molecular and fine structure lines and performing detailed PDR modelling will further elucidate any differences in the ISM conditions.

Ultimately, SPT2349$-$56 provides a key laboratory for probing how efficiently star formation is maintained during the collapse of a protocluster, with potentially important implications for the build-up of the intracluster medium, the early enrichment of the intracluster medium with metals, and the quenching of cluster galaxies at high redshift.

\section*{Acknowledgments}
\begin{acknowledgments}
We would like to thank Desika Narayanan for helpful comments and input. This paper makes use of the following ALMA data: ADS/JAO.ALMA\#2017.1.00273.S, \#2018.1.00018.S, and \#2021.1.01313.S. ALMA is a partnership of ESO (representing its member states), NSF (USA) and NINS (Japan), together with NRC (Canada), MOST and ASIAA (Taiwan), and KASI (Republic of Korea), in cooperation with the Republic of Chile. The Joint ALMA Observatory is operated by ESO, AUI/NRAO and NAOJ. The National Radio Astronomy Observatory is a facility of the National Science Foundation operated under cooperative agreement by Associated Universities, Inc.
S.C. and R.H.\ gratefully acknowledge support for this research from NSERC. MA acknowledges support from ANID Basal Project FB210003 and and ANID MILENIO NCN2024\_112. The Cosmic Dawn Center (DAWN) is funded by the Danish National Research Foundation under grant No. 140. 
\end{acknowledgments}

\bibliography{CI10_bib}{}
\bibliographystyle{aasjournal}

\appendix

\section{Impact of N1 on the protocluster sample statistics} \label{AP:masking_sources}

N1 is the brightest source in SPT2349$-$56 and thus has the most weight when calculating the population statistics, yet this source is certainly not representative of the typical protocluster galaxy. The power law fit parameters and [C{\sc i}] ratio sample statistics calculated excluding N1 are therefore shown in Table \ref{tab:with_outliers} for comparison. Using the mean $R_\mathrm{CI}$ for SPT2349$-$56 calculated without N1, we find $T_{\rm ex}\,{=}\,61.3_{-7.1}^{+8.4}$ K. Excluding N1 therefore does not have a large effect on the population statistics we have calculated.
\begin{table*}
\centering
\caption{Population statistics of the protocluster galaxies in SPT2349$-$56, including and excluding N1.}
\label{tab:with_outliers}
\begin{tabular}{|c|c|c|c|c|c|c|c|c|}
\hline
 \multicolumn{2}{|c|}{} & \multicolumn{4}{c|}{Power Law} & \multicolumn{3}{c|}{\text{C{\sc i}} Ratio}\\
\hline
Including N1 & Line & $\alpha$ & $\gamma$ & Covariance & $\bar{r}$ & $\mu$ & $\sigma$ & $T_{\text{ex}}$\\
\hline
\multirow{2}{*}{Yes} & $\mathrm{[\text{C{\sc i}}](1\text{--}0)}$ & 1.243$\pm$0.076 & 0.656$\pm$0.093 & -0.00069 & 1.41$\pm$0.26 & \multirow{2}{*}{1.094$\pm$0.090} & \multirow{2}{*}{0.42$\pm$0.11} & \multirow{2}{*}{$59.1_{-6.8}^{+8.1}$\,K}\\ \cline{2-6}

& $\mathrm{[\text{C{\sc i}}](2\text{--}1)}$ & 5.57$\pm$0.37 & 0.729$\pm$0.094 & 0.015 & 2.17$\pm$0.38 & & &\\ \hline

\multirow{2}{*}{No} & $\mathrm{[\text{C{\sc i}}](1\text{--}0)}$ & 1.122$\pm$0.089 & 0.48$\pm$0.10 & 0.0047 & 1.34$\pm$0.26 & \multirow{2}{*}{1.120$\pm$0.099} & \multirow{2}{*}{0.43$\pm$0.12} & \multirow{2}{*}{$61.3_{-7.1}^{+8.4}$\,K}\\ \cline{2-6}

& $\mathrm{[\text{C{\sc i}}](2\text{--}1)}$ & 5.96$\pm$0.48 & 0.84$\pm$0.13 & 0.041 & 2.24$\pm$0.38 & & &\\ \hline
\end{tabular}
\end{table*}

\section{Spatially-coincident [C{\sc i}] and [C{\sc ii}] line emission cutouts} \label{AP:line_cutouts}

Here we provide [C{\sc i}](1--0), [C{\sc i}](2--1), and [C{\sc ii}] cutouts of the protocluster galaxies in SPT2349$-$56 after averaging over the integration range used in our line detection pipeline. Fig.~\ref{fig:cutouts} shows [C{\sc ii}] cutouts and contours (blue) in steps of $2^n \sigma$, where $n\,{=}\,0,1,2,3...$, with [C{\sc i}](1--0) (red) and [C{\sc i}](2--1) (yellow) contours overlaid at the same $\sigma$ levels for comparison. For all lines, the integration range was set to $2\sigma$ around the center of the line. The peak pixel in the [C{\sc ii}] cutouts are indicated by circles, and we find that for most of our ${>}\,2.5\sigma$ detections, the [C{\sc ii}] falls within the 2$\sigma$ contours of the [C{\sc i}](1--0) and [C{\sc i}](2--1) maps.

\begin{figure*}
    \centering
    \fbox{
    \parbox{0.3\textwidth}{
    \includegraphics[width = 0.3\textwidth]{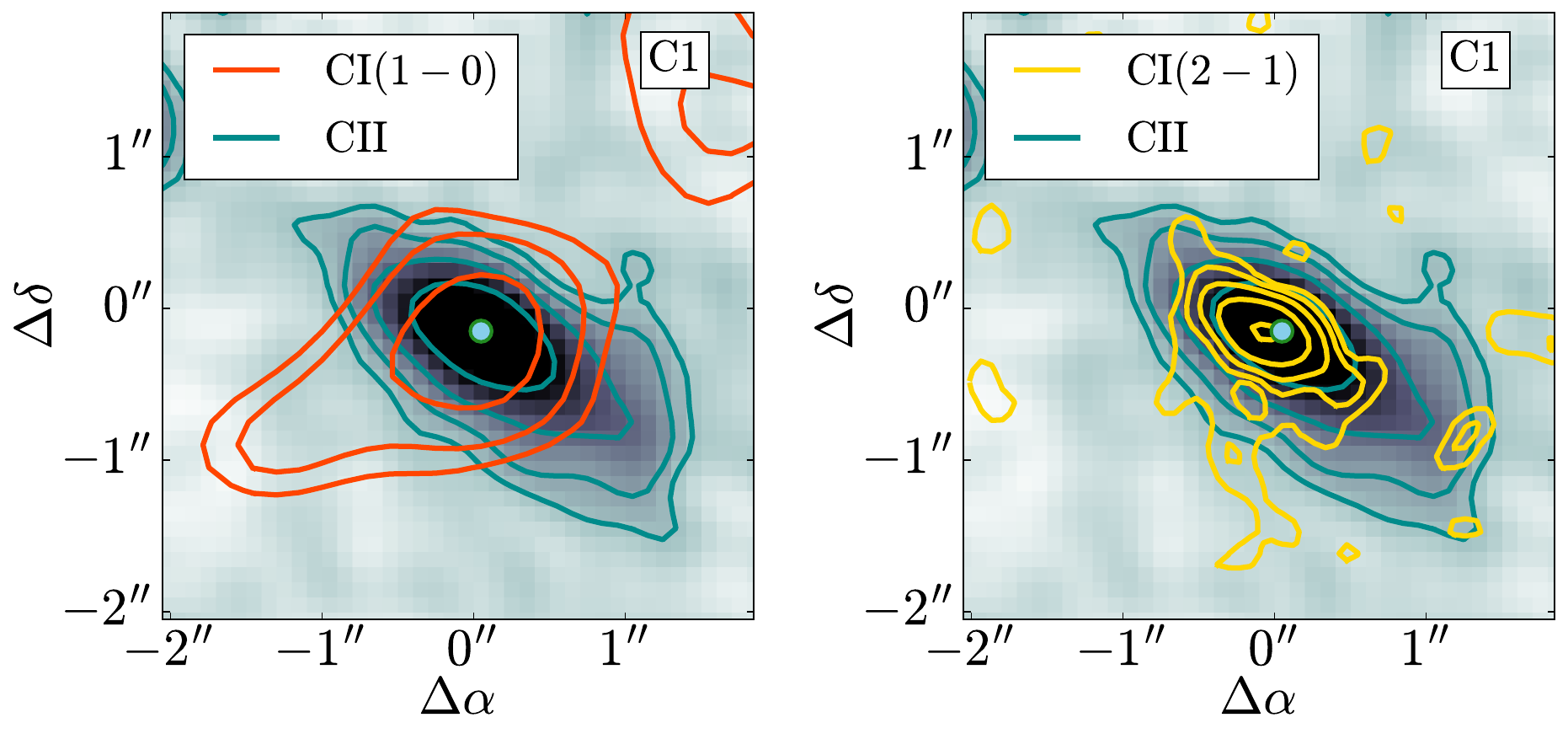}}}
    \fbox{
    \parbox{0.3\textwidth}{
    \includegraphics[width = 0.3\textwidth]{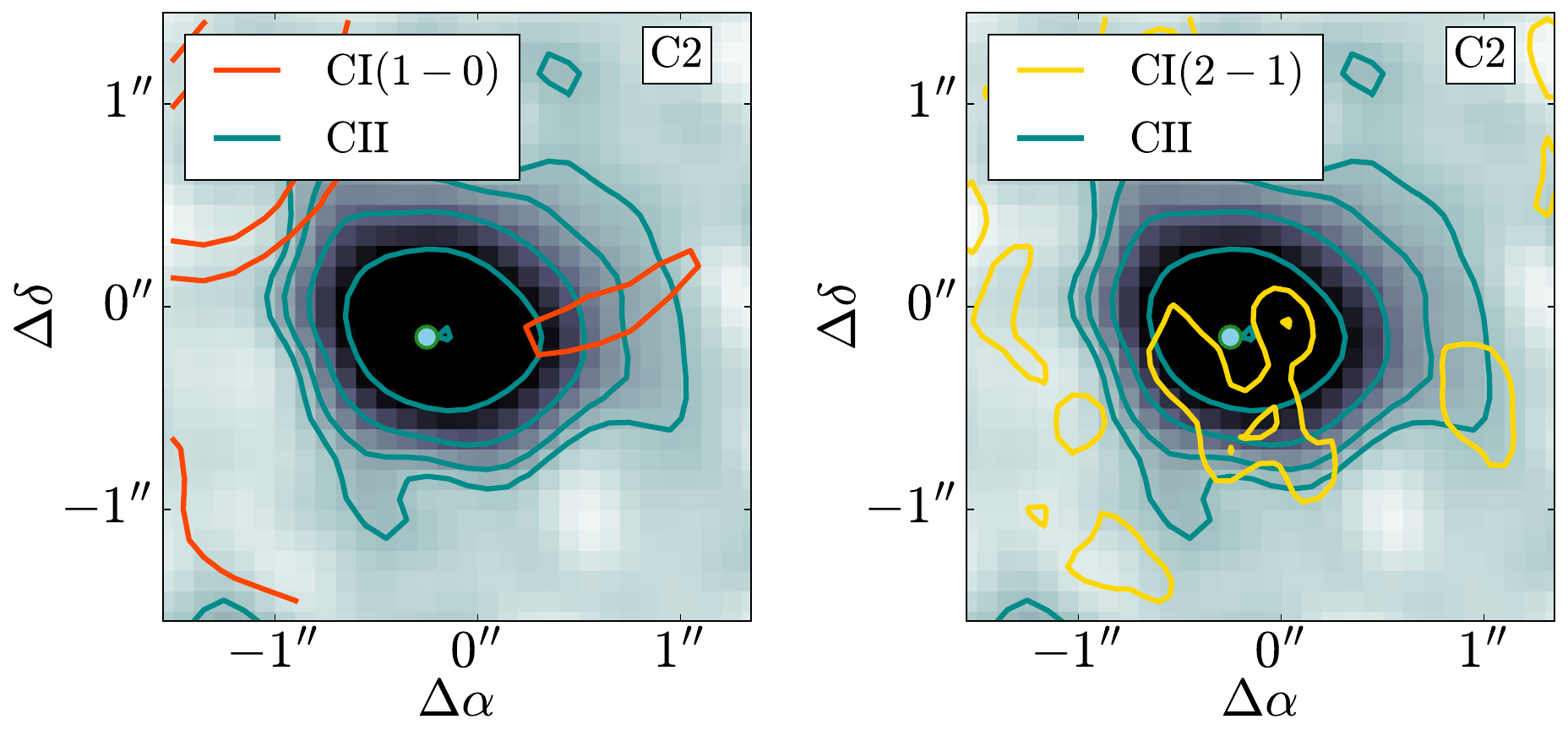}}}
    \fbox{
    \parbox{0.3\textwidth}{
    \includegraphics[width = 0.3\textwidth]{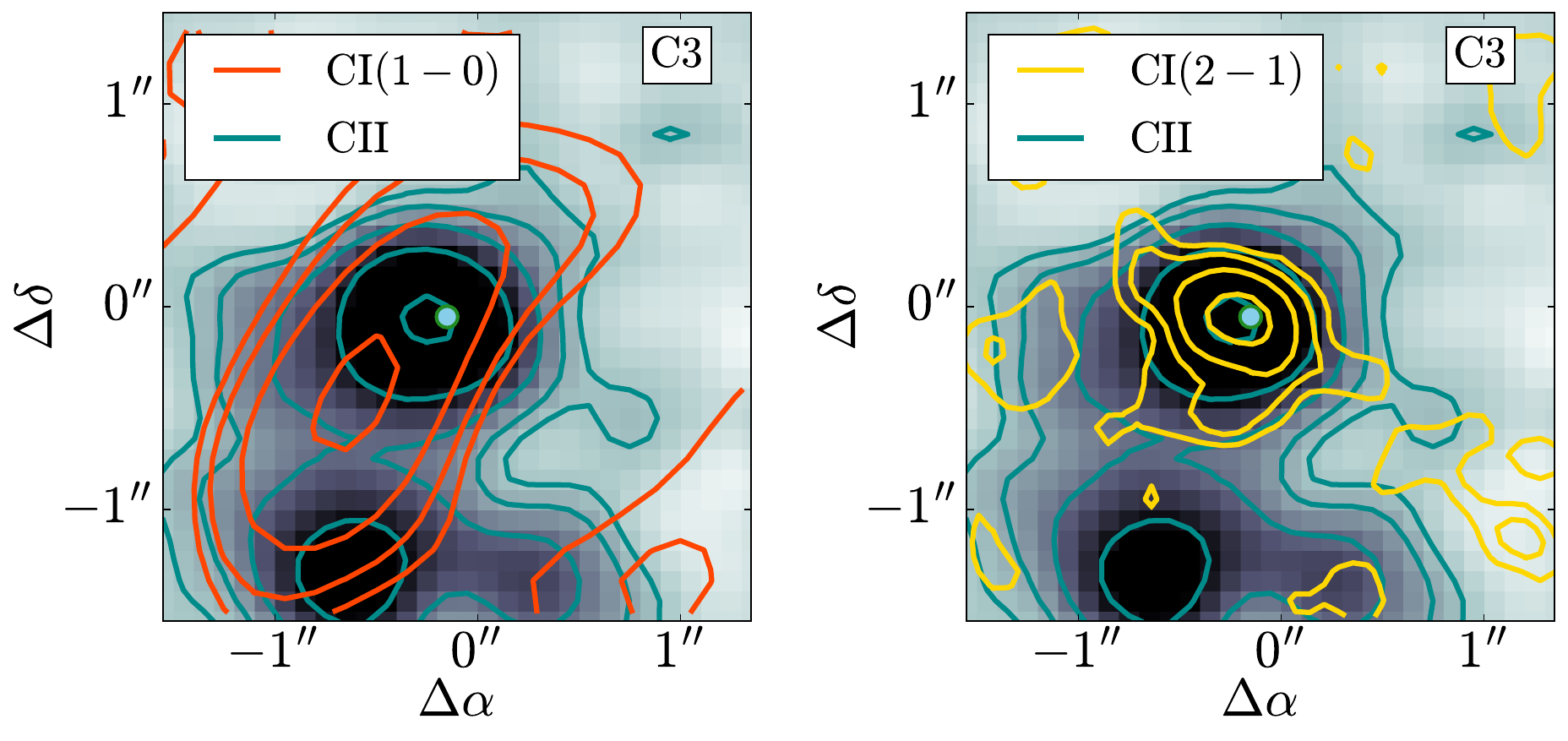}}}
    \fbox{
    \parbox{0.3\textwidth}{
    \includegraphics[width = 0.3\textwidth]{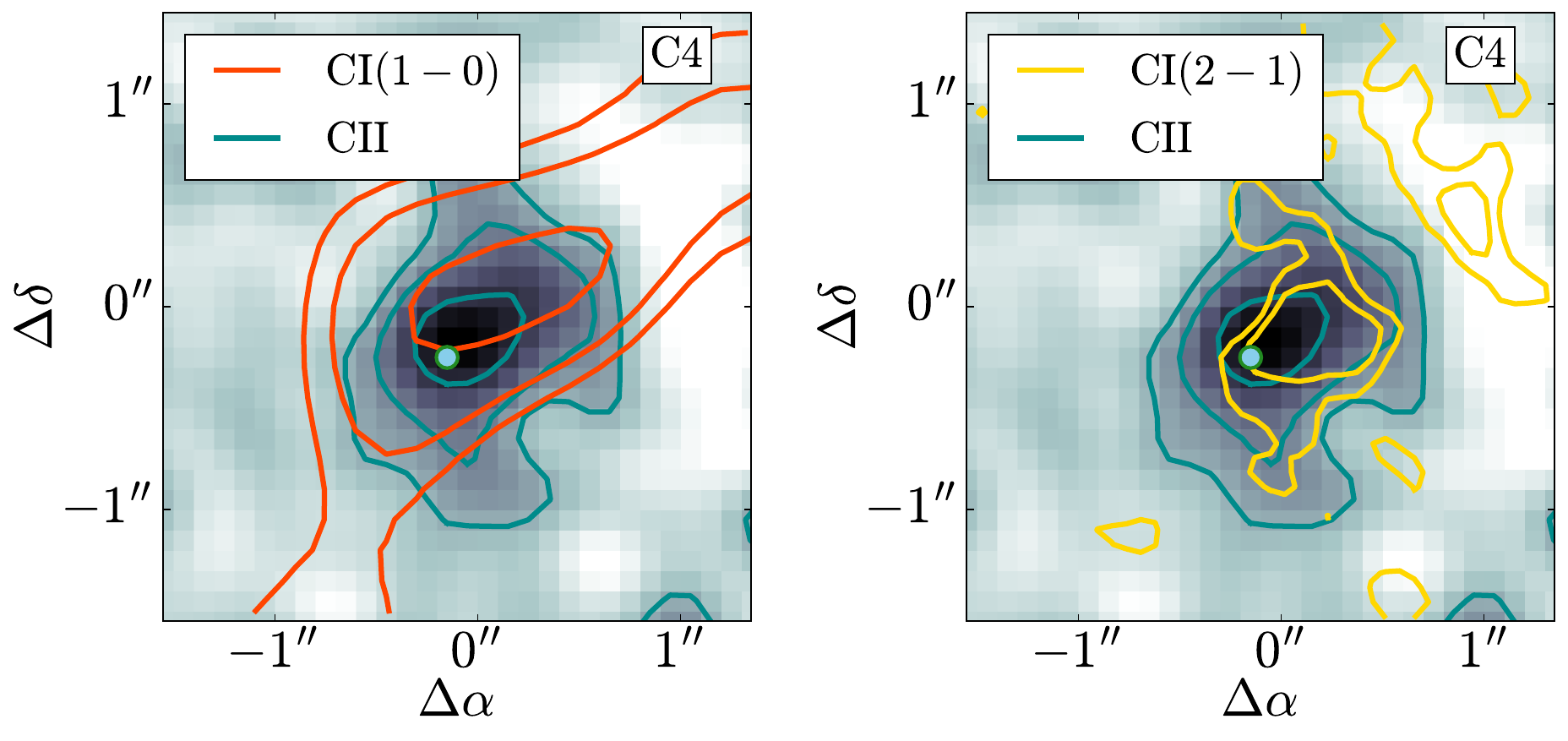}}}
    \fbox{
    \parbox{0.3\textwidth}{
    \includegraphics[width = 0.3\textwidth]{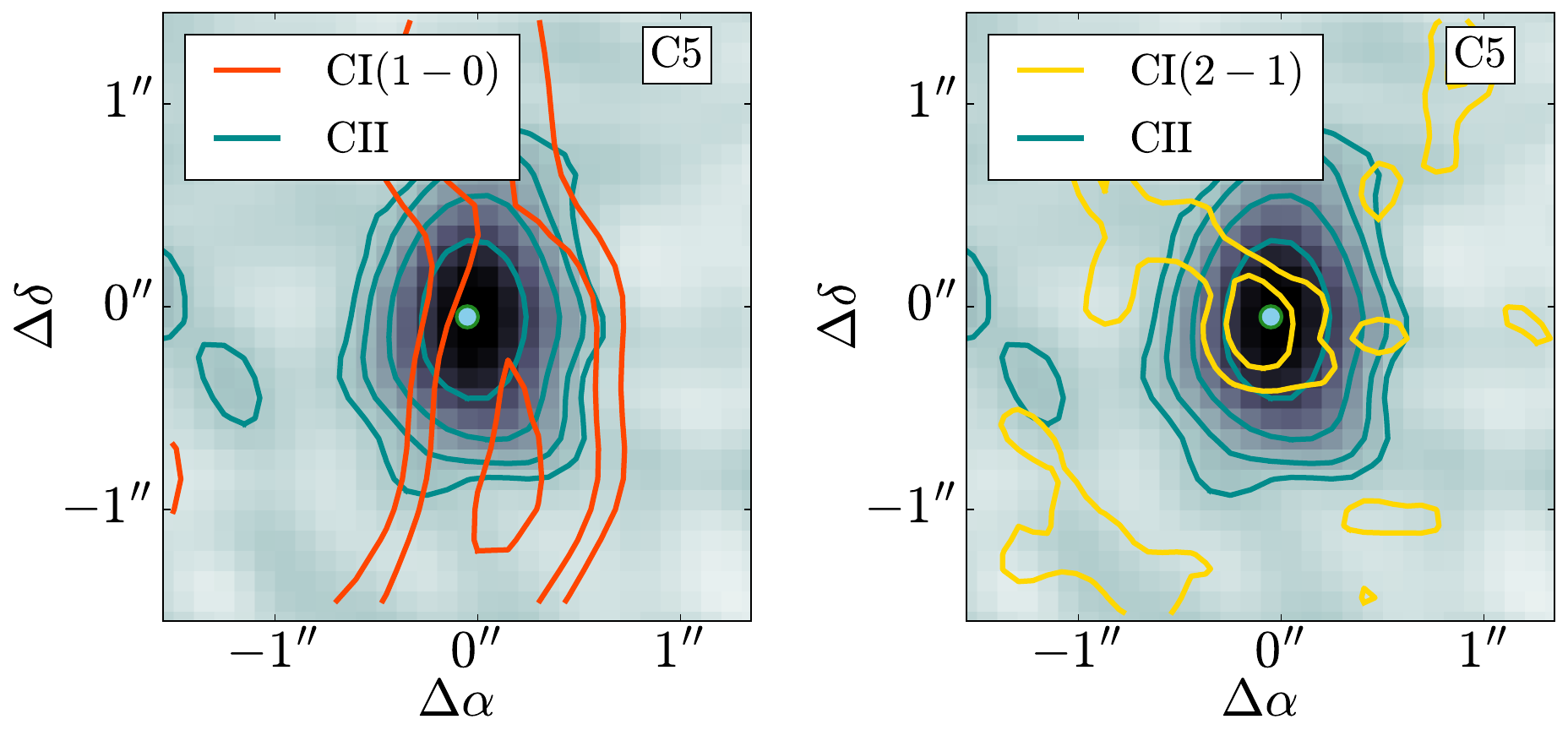}}}
    \fbox{
    \parbox{0.3\textwidth}{
    \includegraphics[width = 0.3\textwidth]{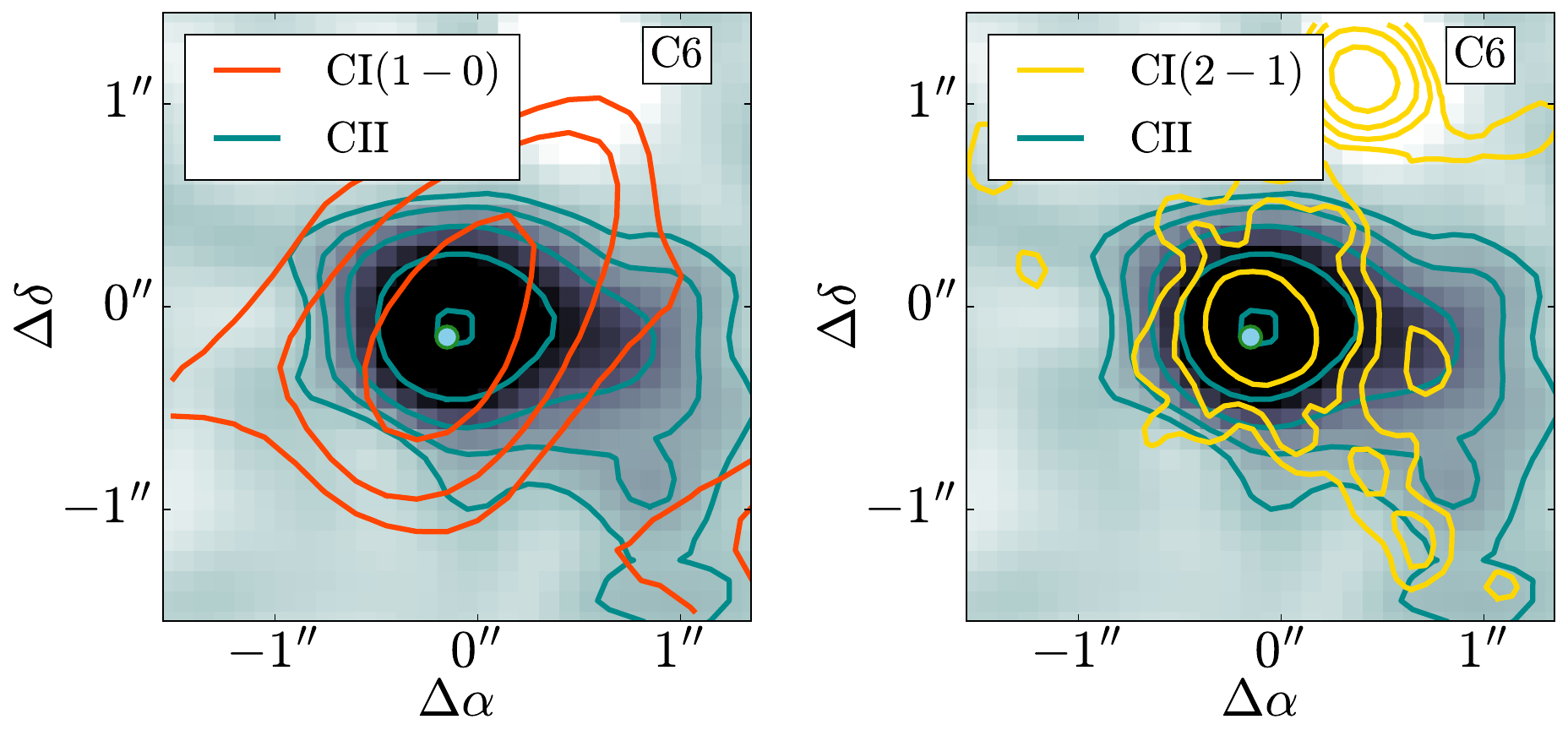}}}
    \fbox{
    \parbox{0.3\textwidth}{
    \includegraphics[width = 0.3\textwidth]{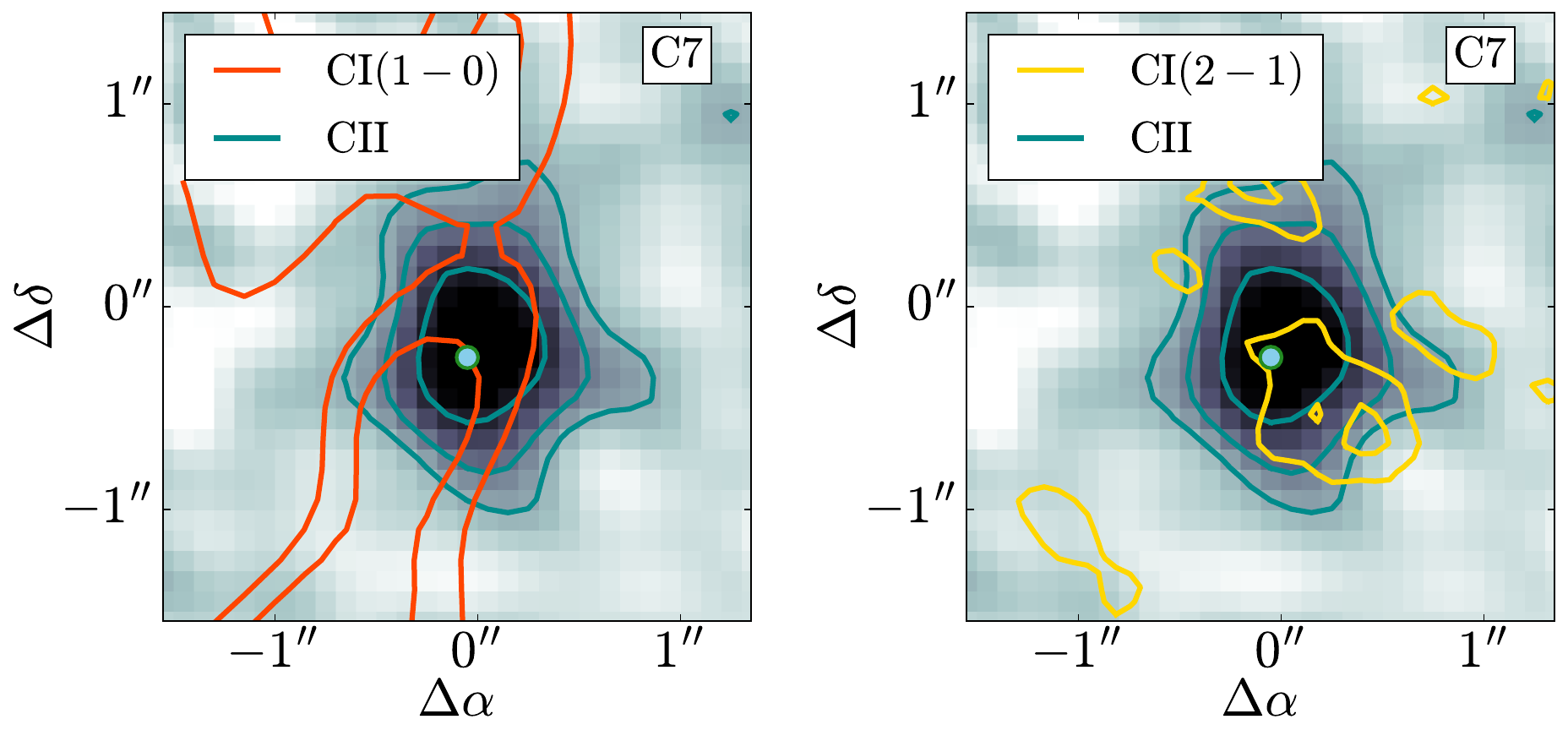}}}
    \fbox{
    \parbox{0.3\textwidth}{
    \includegraphics[width = 0.3\textwidth]{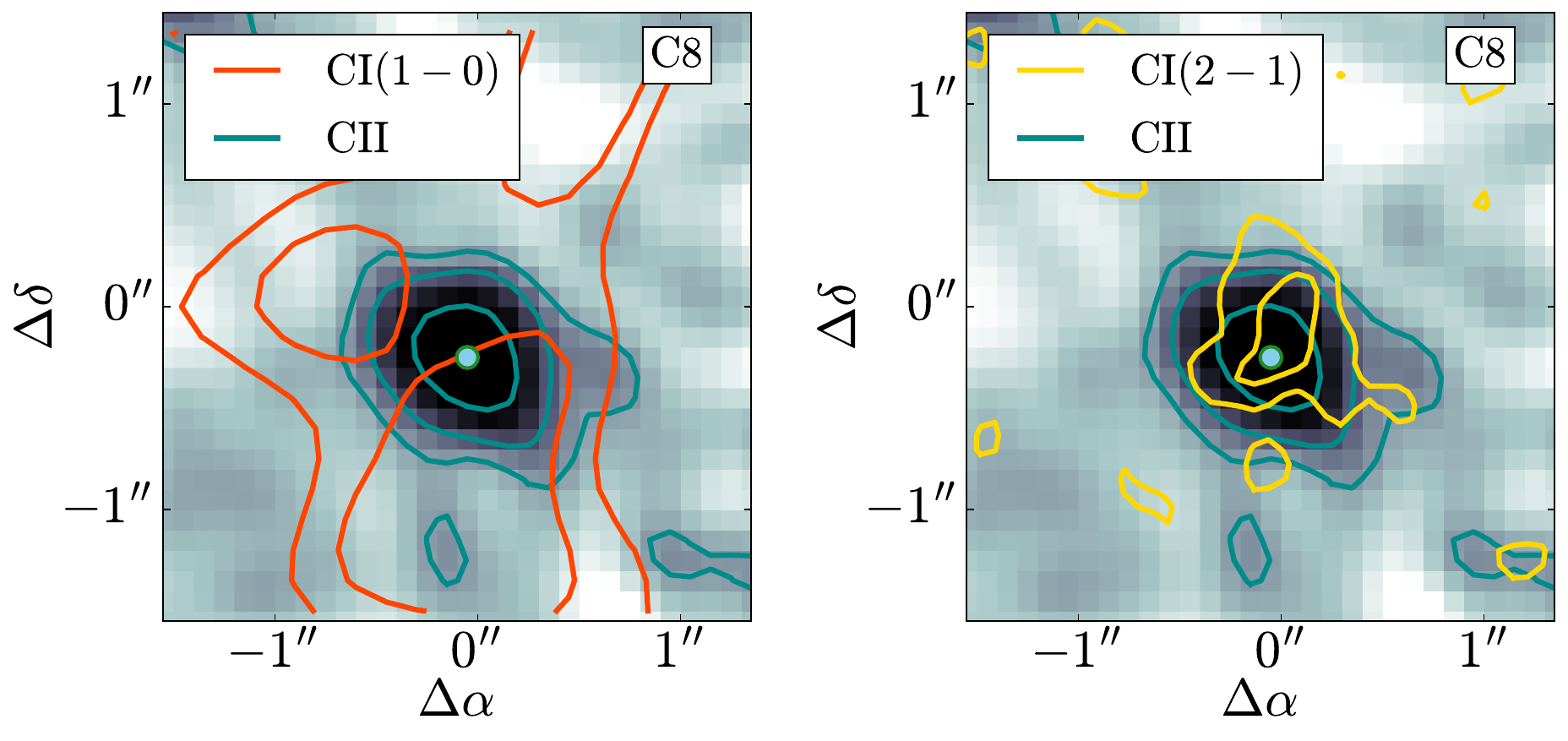}}}
    \fbox{
    \parbox{0.3\textwidth}{
    \includegraphics[width = 0.3\textwidth]{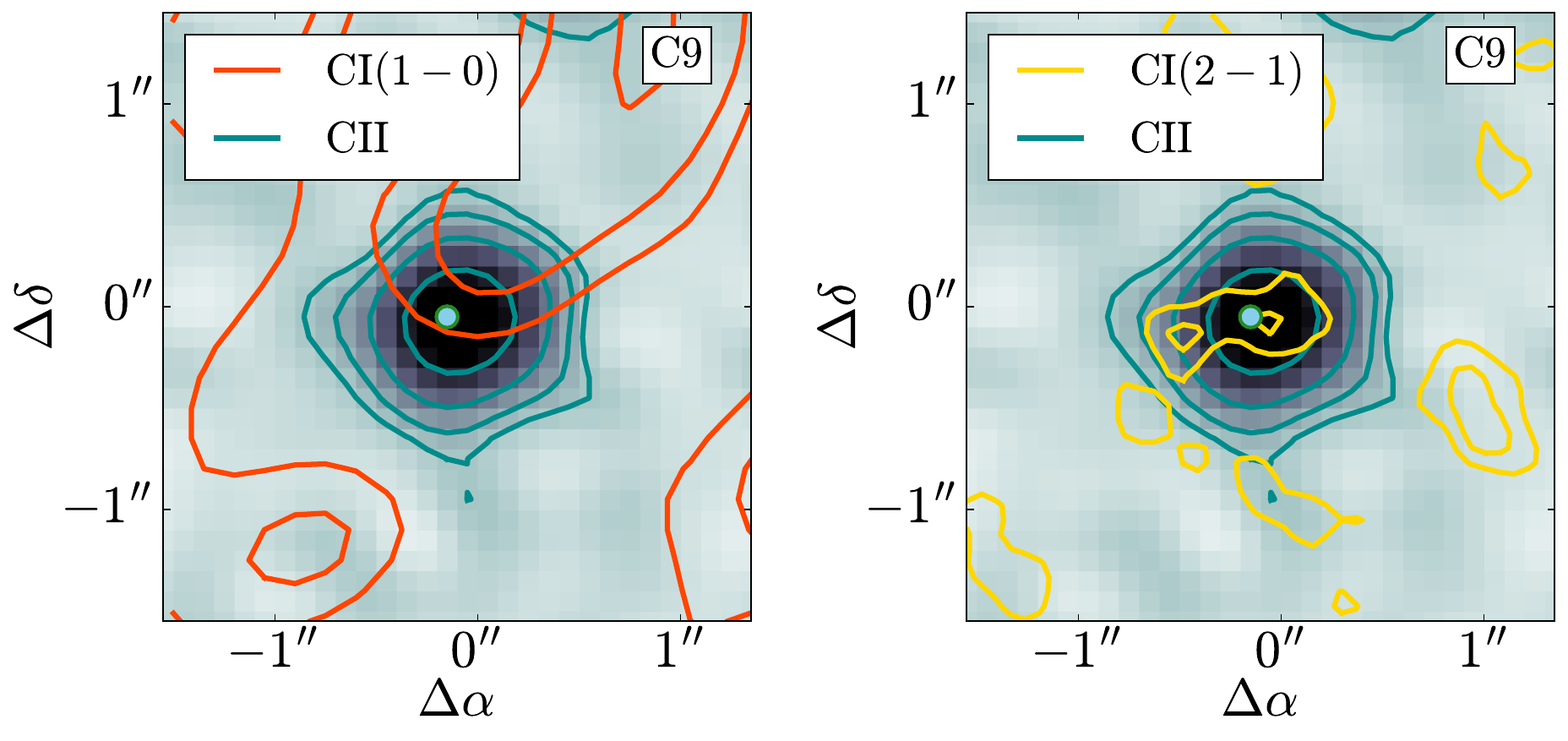}}}
    \fbox{
    \parbox{0.3\textwidth}{
    \includegraphics[width = 0.3\textwidth]{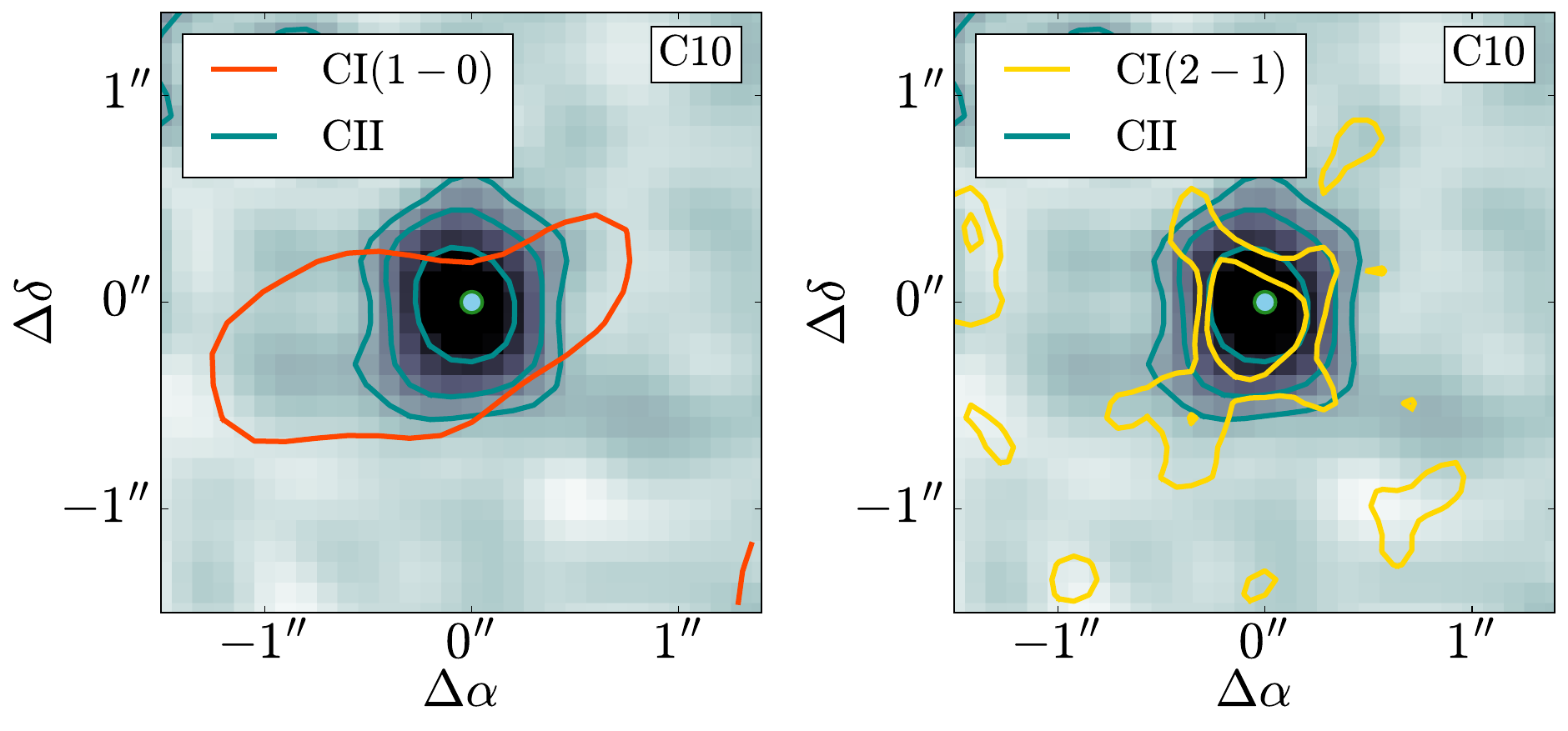}}}
    \fbox{
    \parbox{0.3\textwidth}{
    \includegraphics[width = 0.3\textwidth]{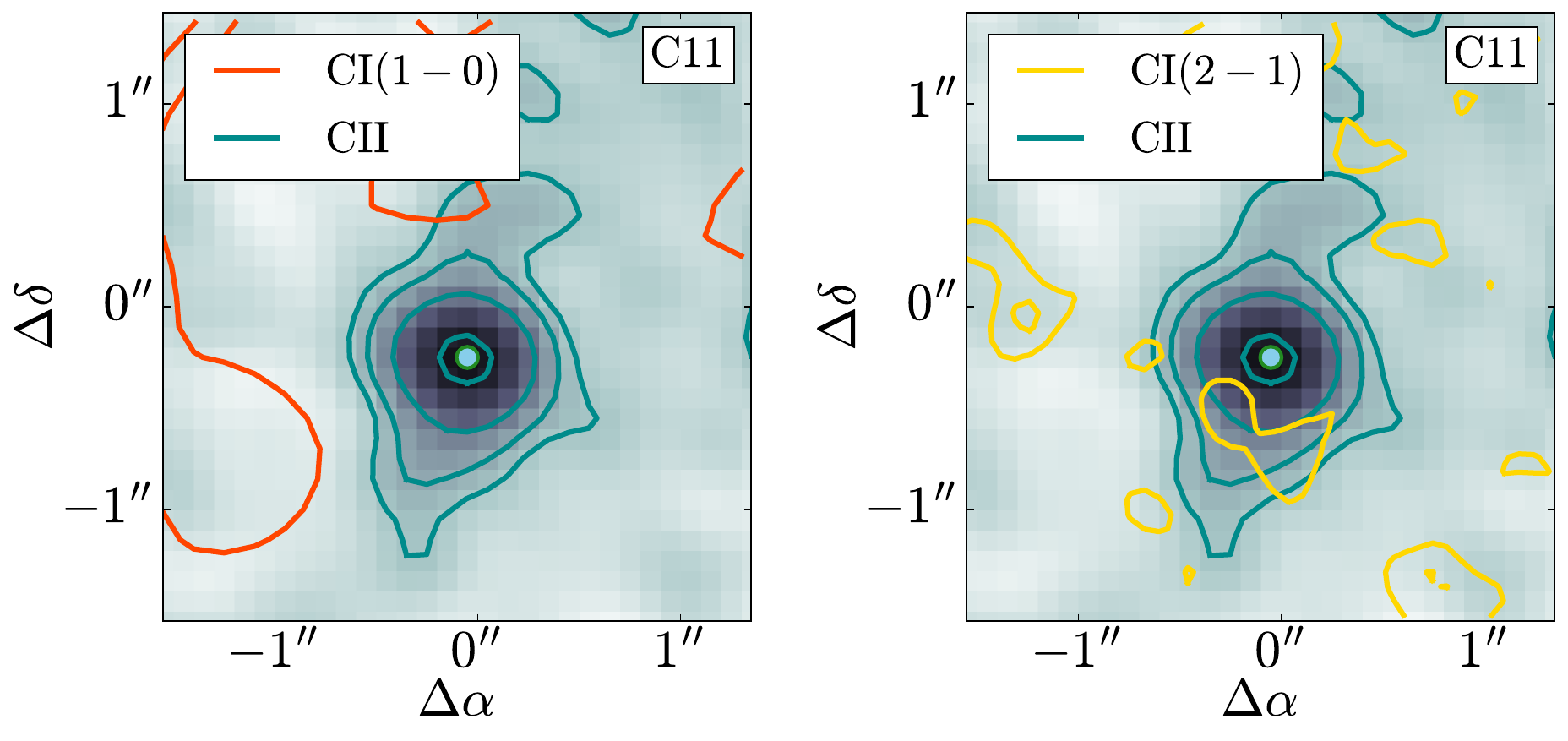}}}
    \fbox{
    \parbox{0.3\textwidth}{
    \includegraphics[width = 0.3\textwidth]{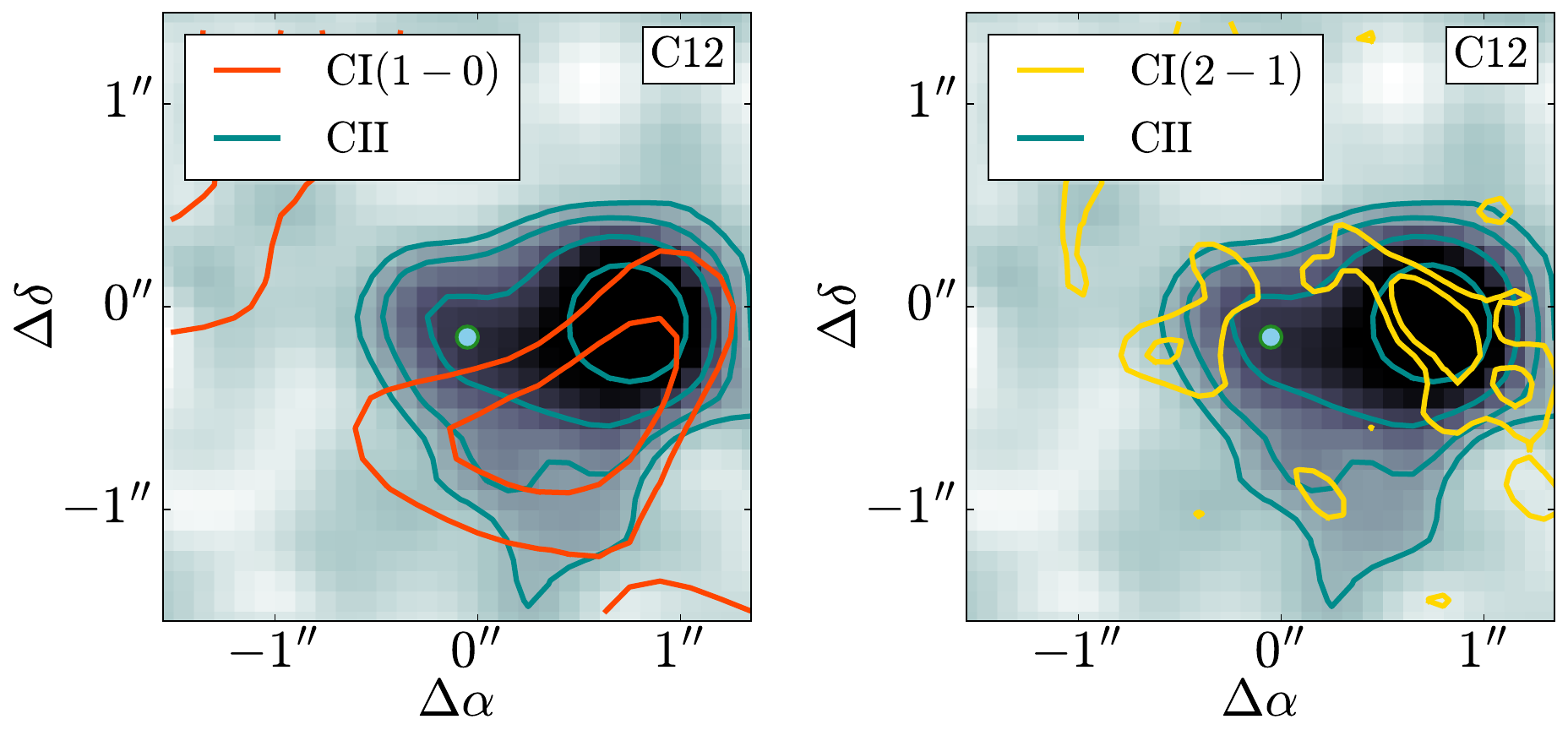}}}
    \fbox{
    \parbox{0.3\textwidth}{
    \includegraphics[width = 0.3\textwidth]{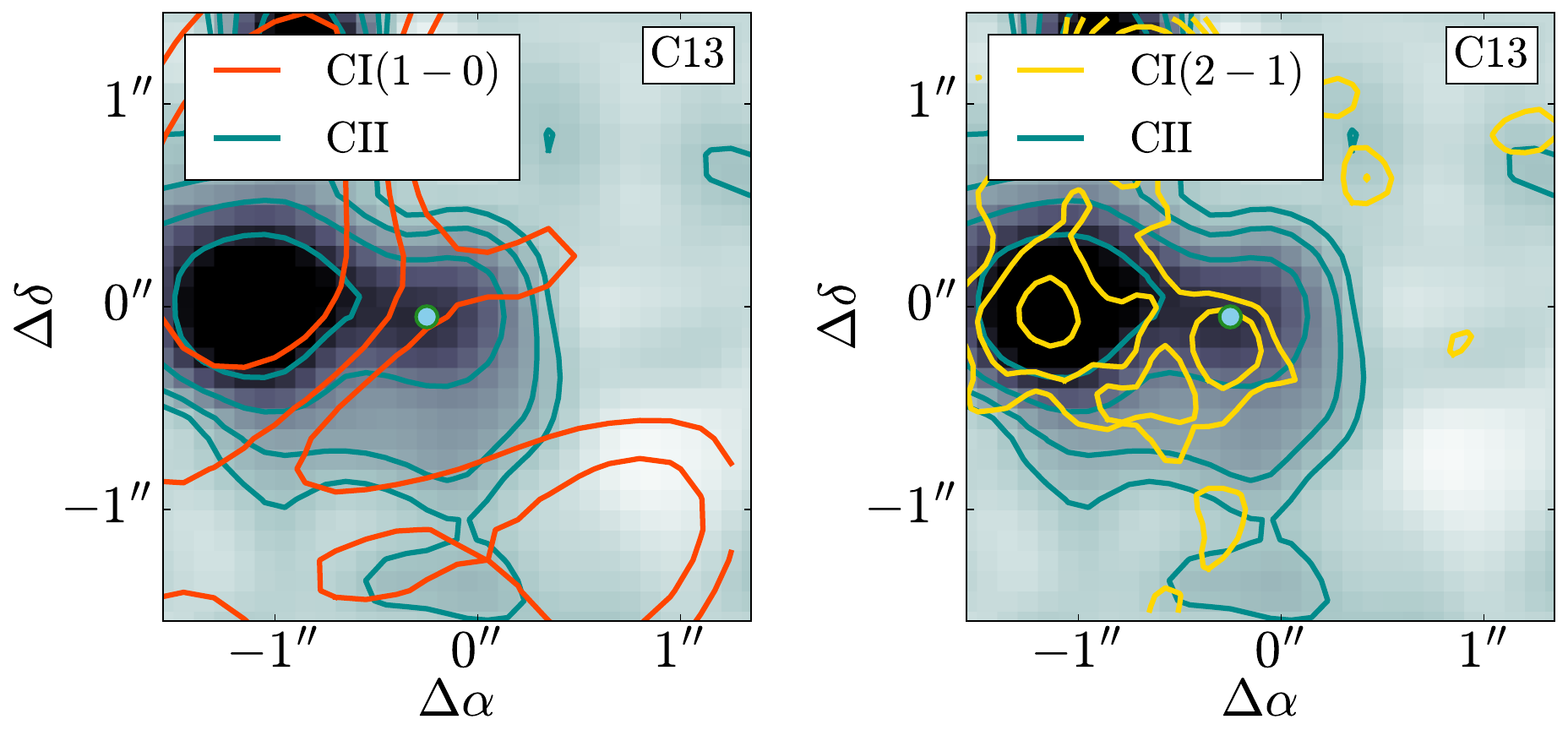}}}
    \fbox{
    \parbox{0.3\textwidth}{
    \includegraphics[width = 0.3\textwidth]{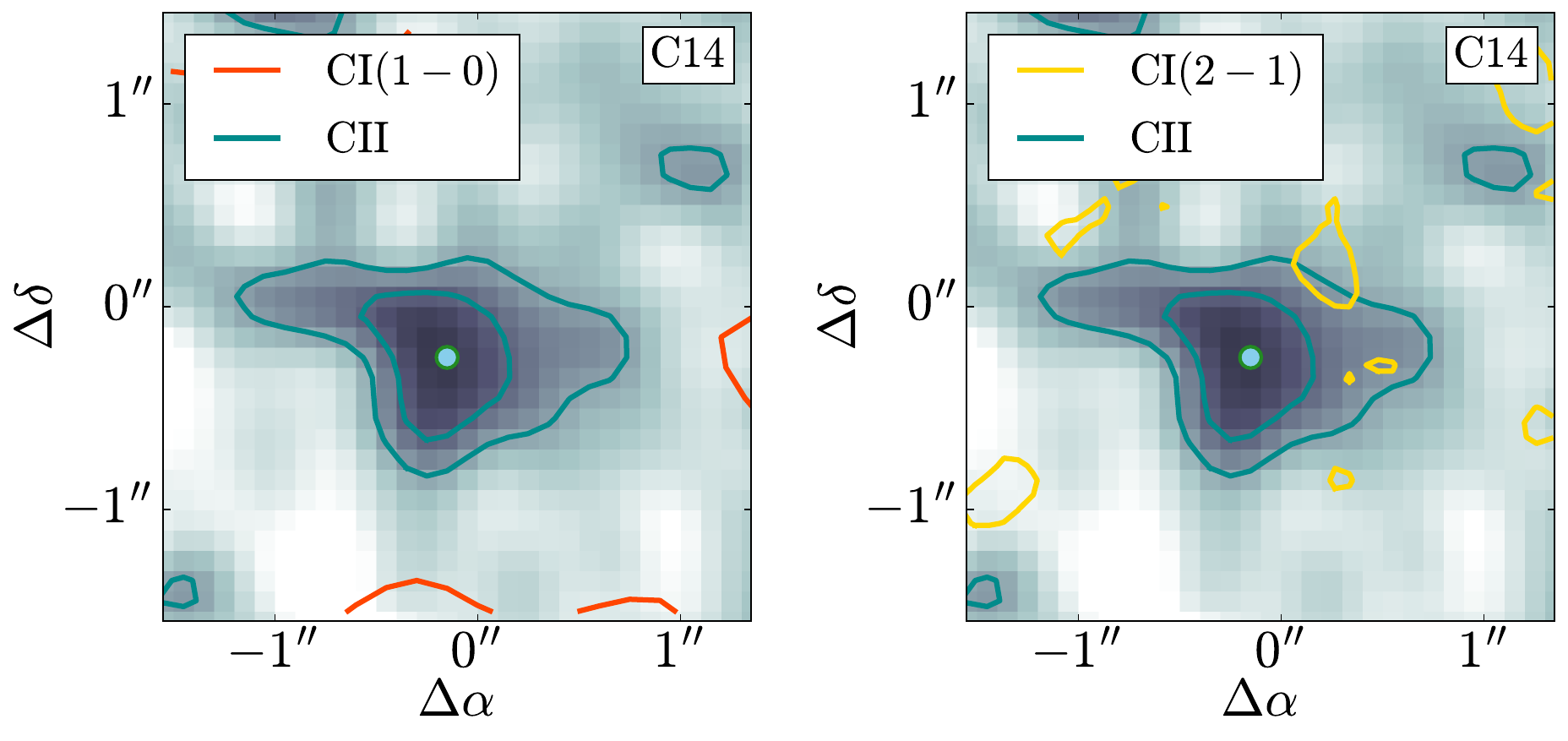}}}
    \fbox{
    \parbox{0.3\textwidth}{
    \includegraphics[width = 0.3\textwidth]{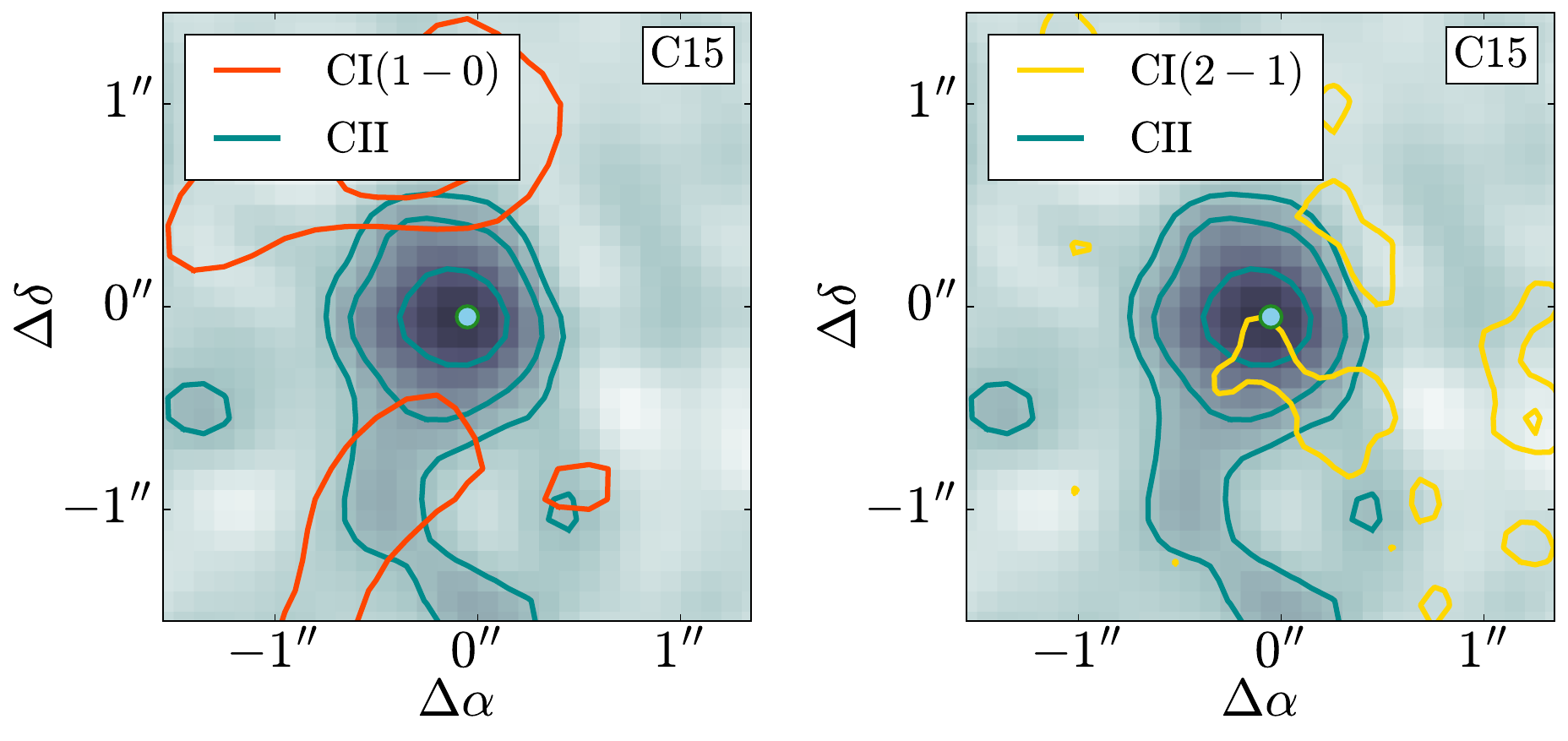}}}
    \fbox{
    \parbox{0.3\textwidth}{
    \includegraphics[width = 0.3\textwidth]{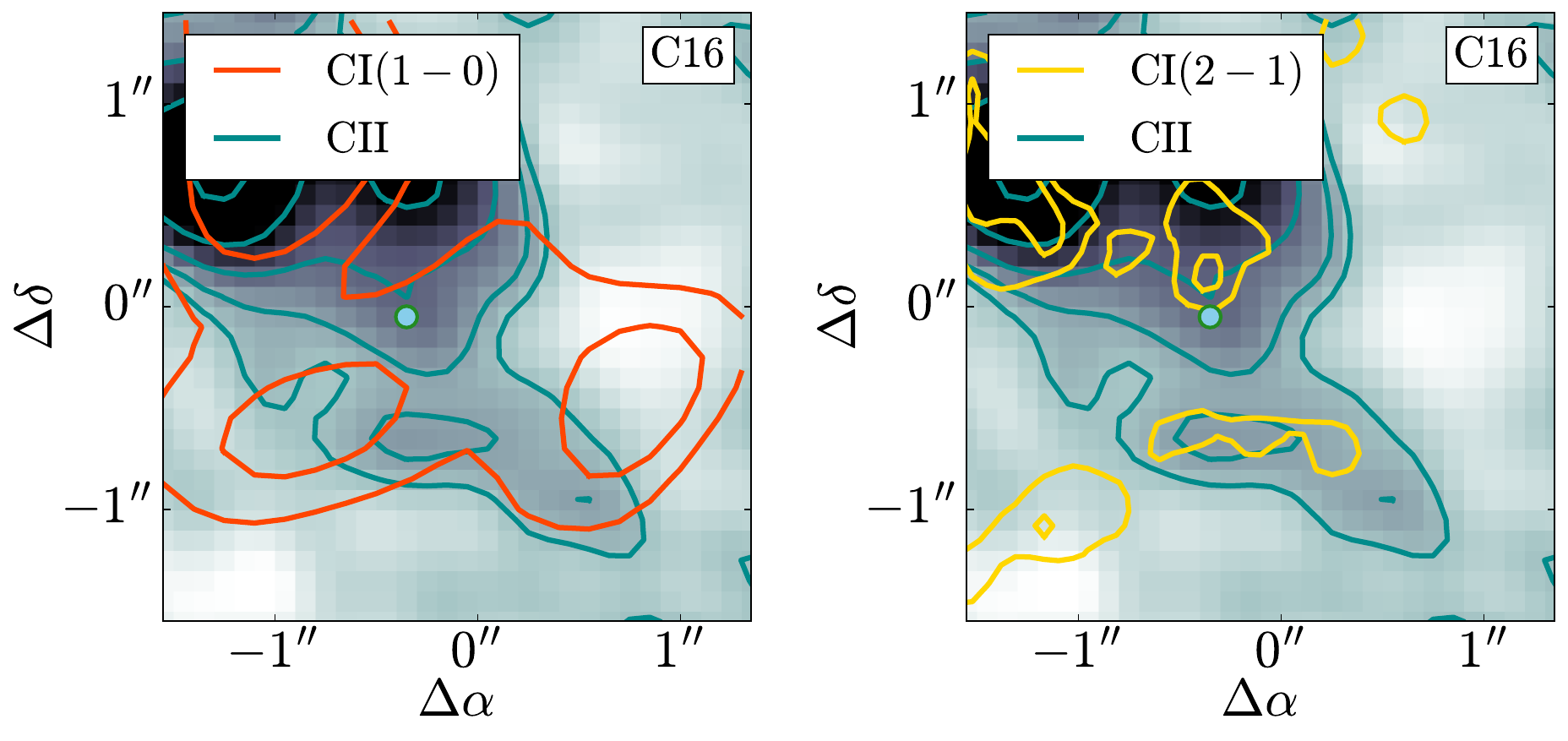}}}
    \fbox{
    \parbox{0.3\textwidth}{
    \includegraphics[width = 0.3\textwidth]{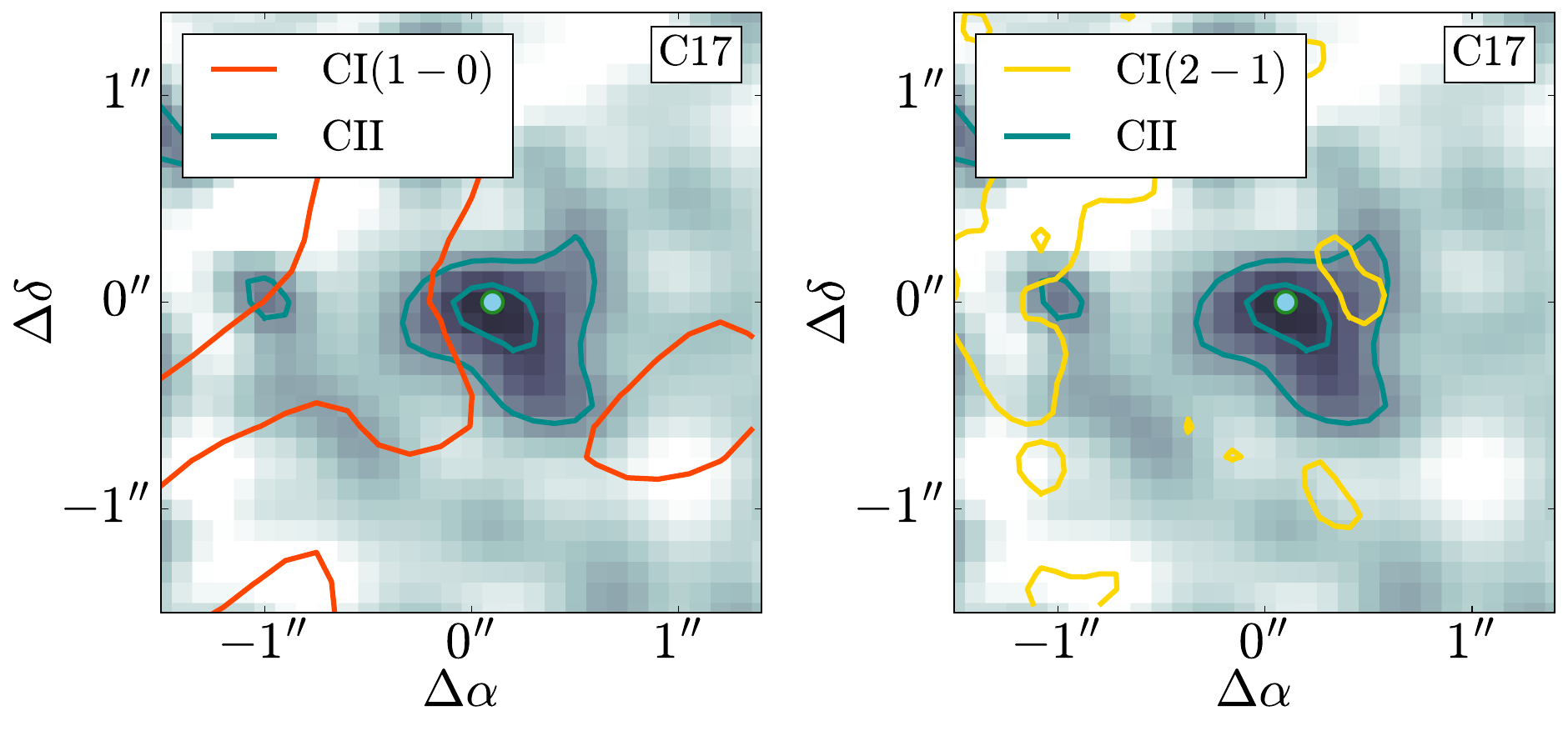}}}
    \fbox{
    \parbox{0.3\textwidth}{
    \includegraphics[width = 0.3\textwidth]{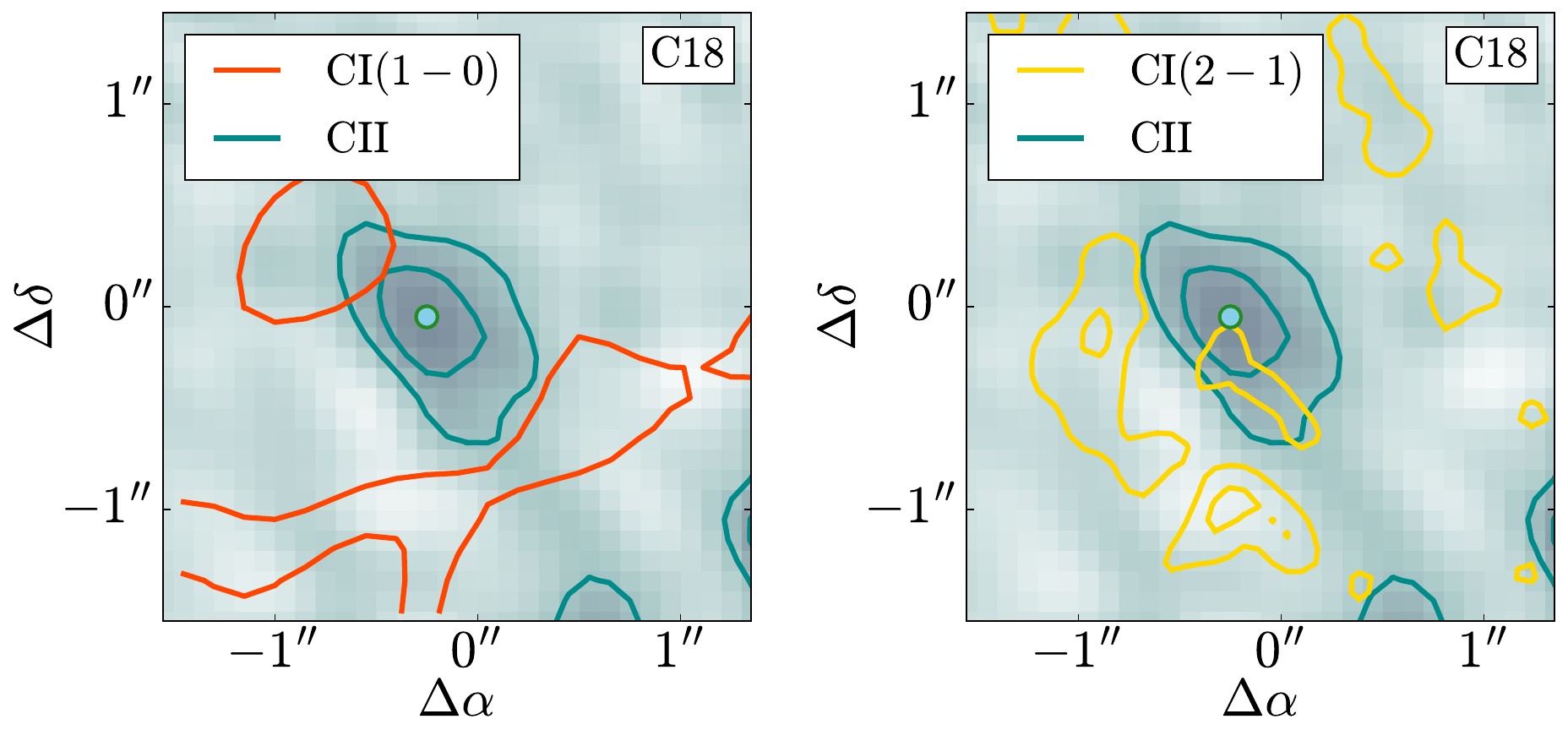}}}
    \fbox{
    \parbox{0.3\textwidth}{
    \includegraphics[width = 0.3\textwidth]{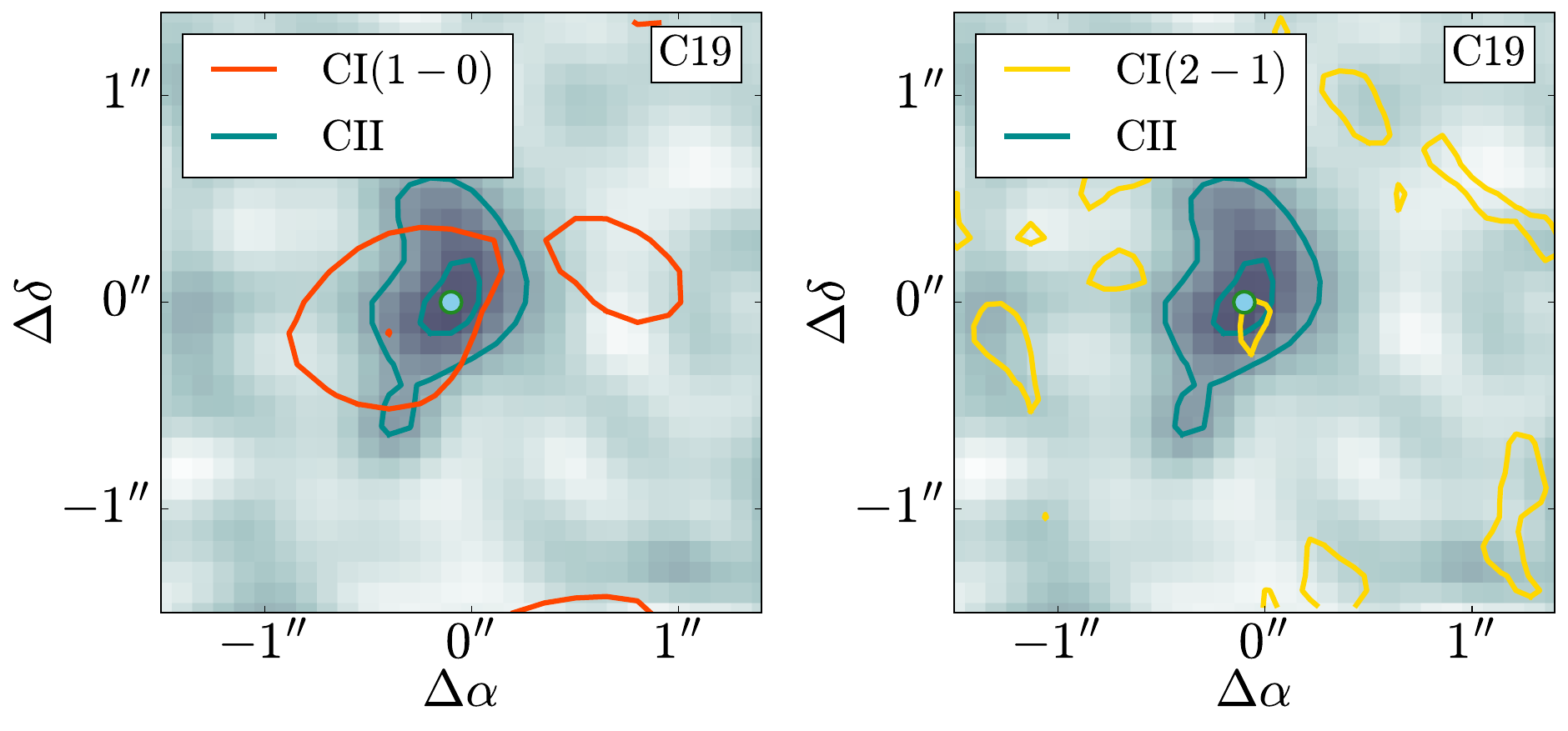}}}
    \fbox{
    \parbox{0.3\textwidth}{
    \includegraphics[width = 0.3\textwidth]{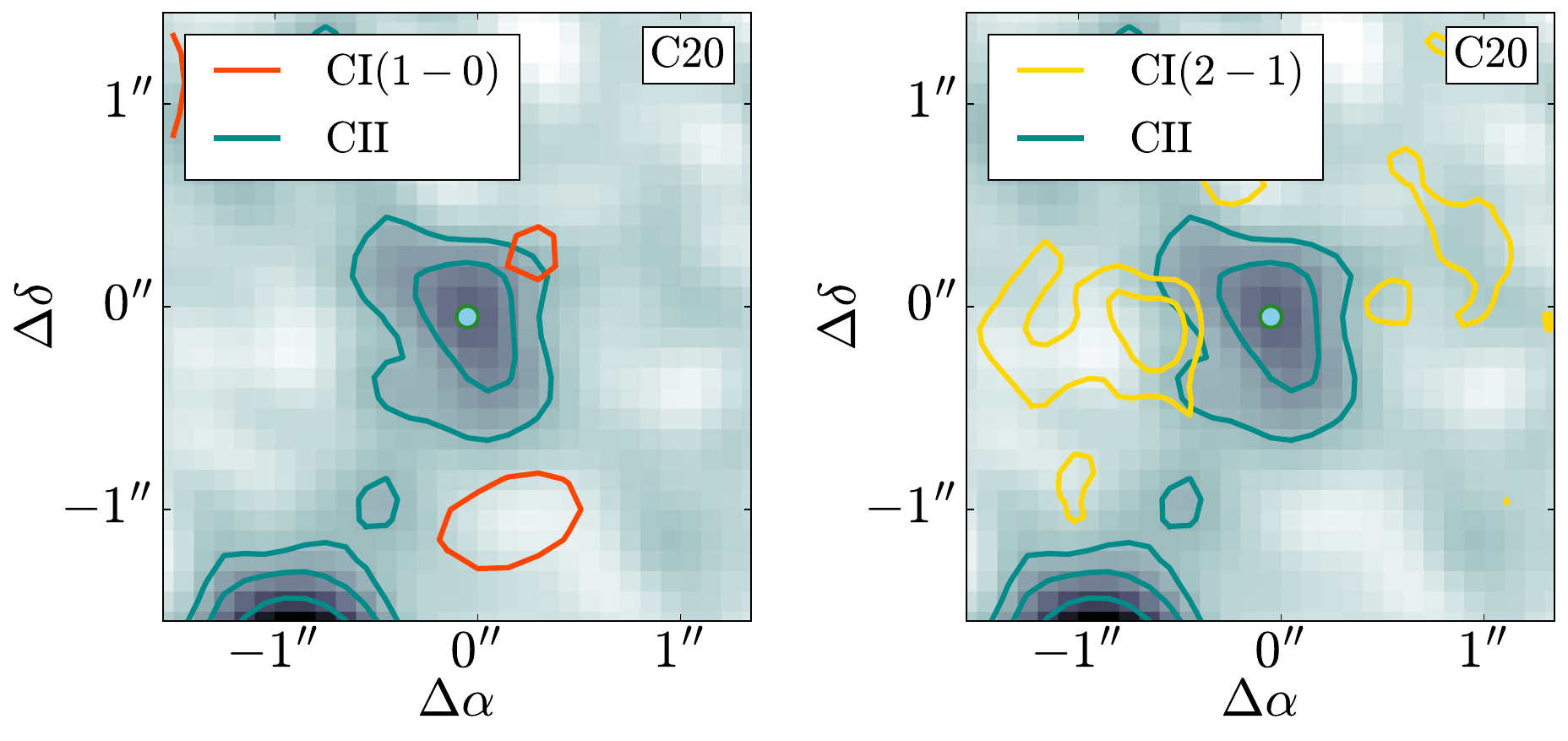}}}
    \fbox{
    \parbox{0.3\textwidth}{
    \includegraphics[width = 0.3\textwidth]{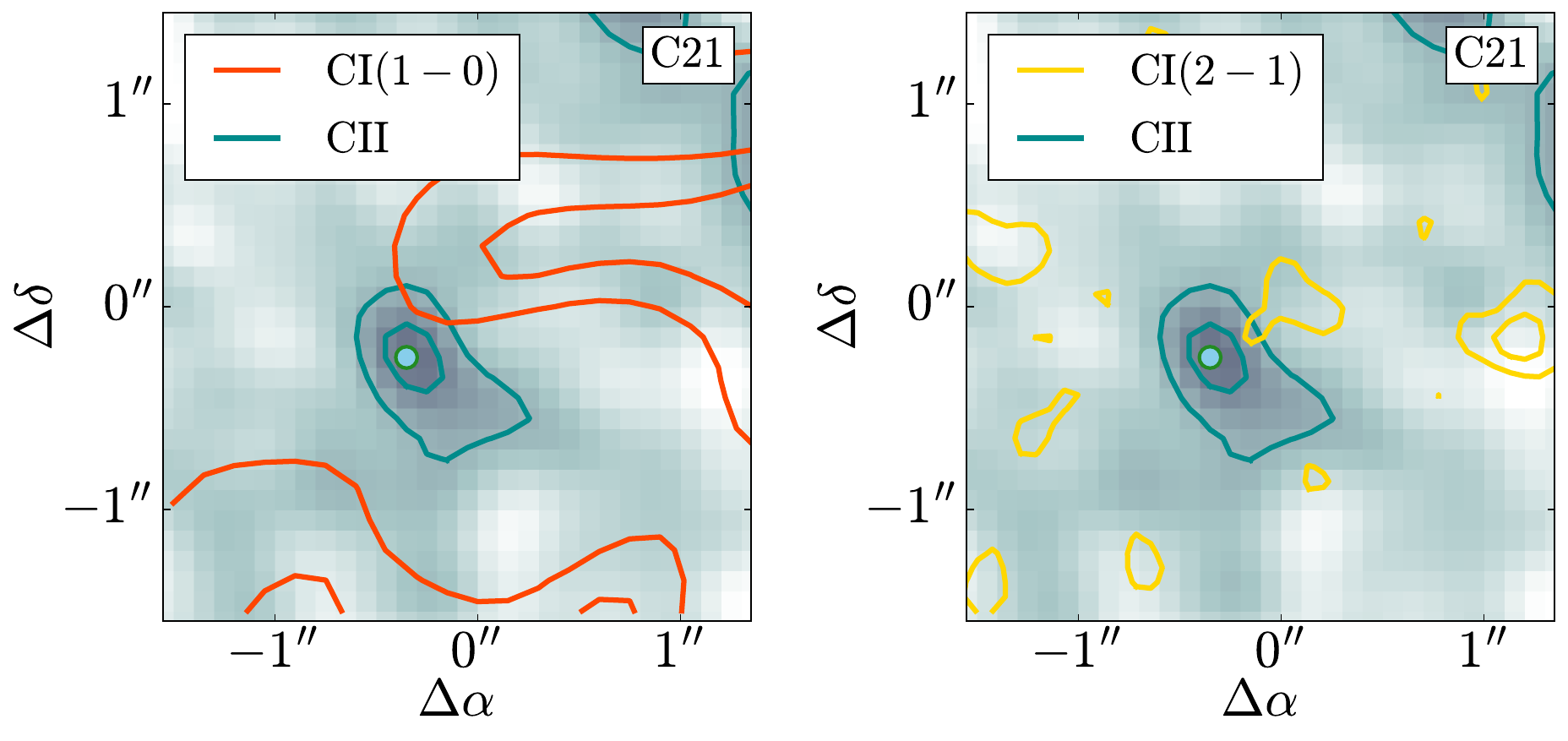}}}
    \caption{[C{\sc ii}] cutouts from \citet{hill2020} are shown in the background, with blue contours overlaid in steps of $2^n \sigma$ (where $n\,{=}\,0,1,2,3...$). [C{\sc i}](1--0) (red) and [C{\sc i}](2--1) (yellow) contours are overlaid following the same $\sigma$ levels for comparison. The peak pixel in the [C{\sc ii}] cutouts are indicated by circles. In all cases the line integration ranges are $2\sigma$ around the centers of the lines. The [C{\sc i}](2--1) maps have been lightly smoothed by a Gaussian with a standard deviation of 1 pixel for presentation purposes.} 
    \label{fig:cutouts}
\end{figure*}

\begin{figure*}
    \setcounter{figure}{3}
    \centering
    \fbox{
    \parbox{0.3\textwidth}{
    \includegraphics[width = 0.3\textwidth]{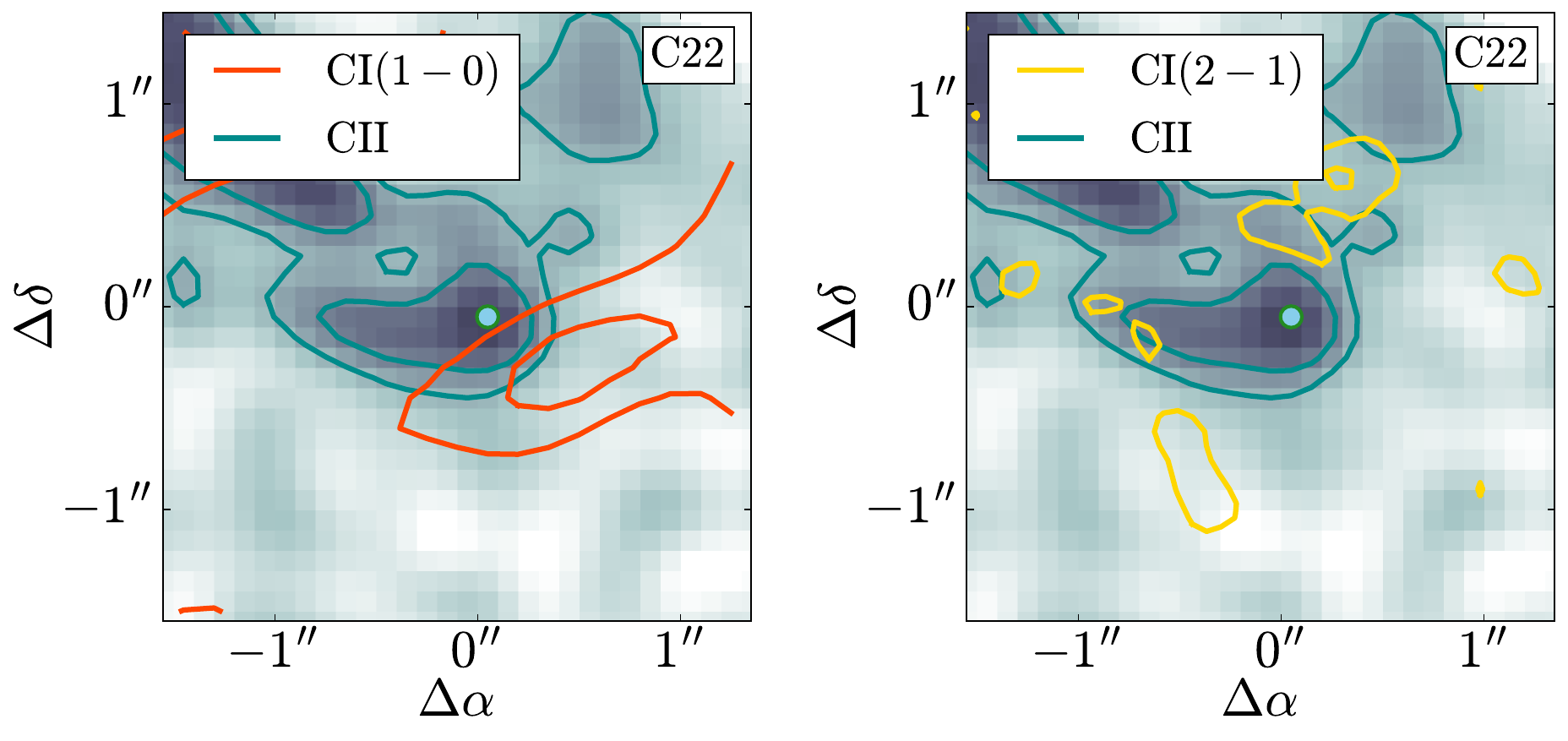}}}
    \fbox{
    \parbox{0.3\textwidth}{
    \includegraphics[width = 0.3\textwidth]{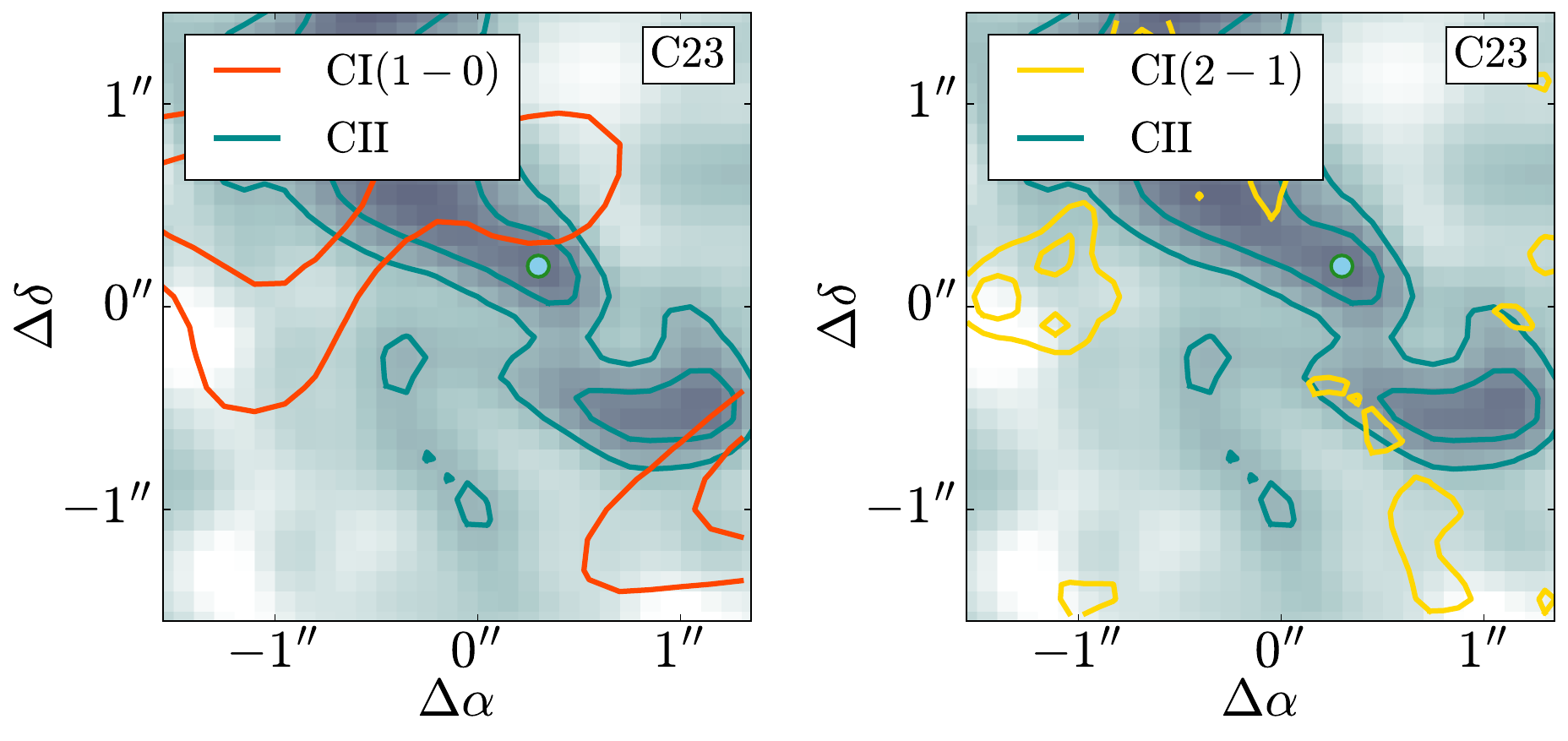}}}
    \fbox{
    \parbox{0.3\textwidth}{
    \includegraphics[width = 0.3\textwidth]{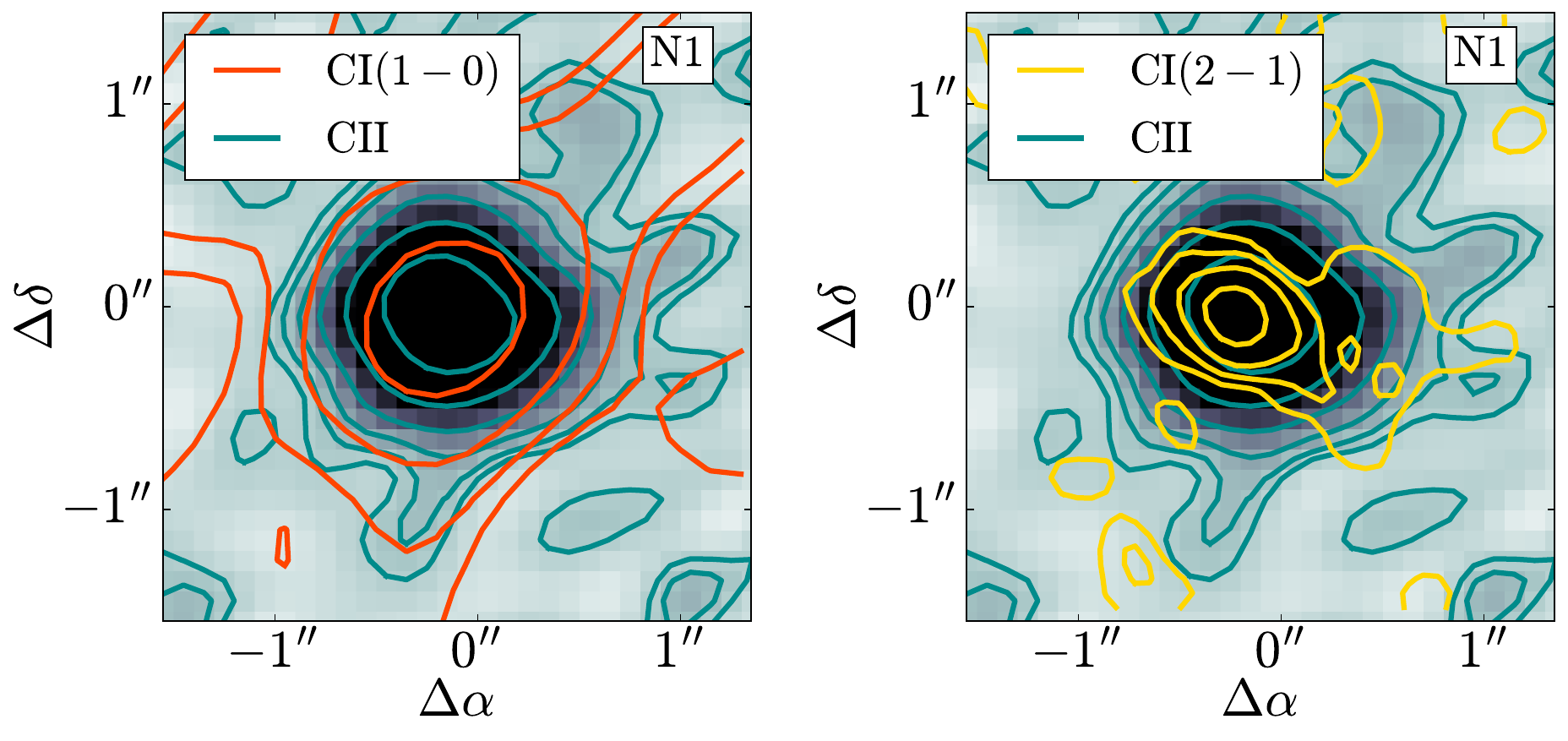}}}
    \fbox{
    \parbox{0.3\textwidth}{
    \includegraphics[width = 0.3\textwidth]{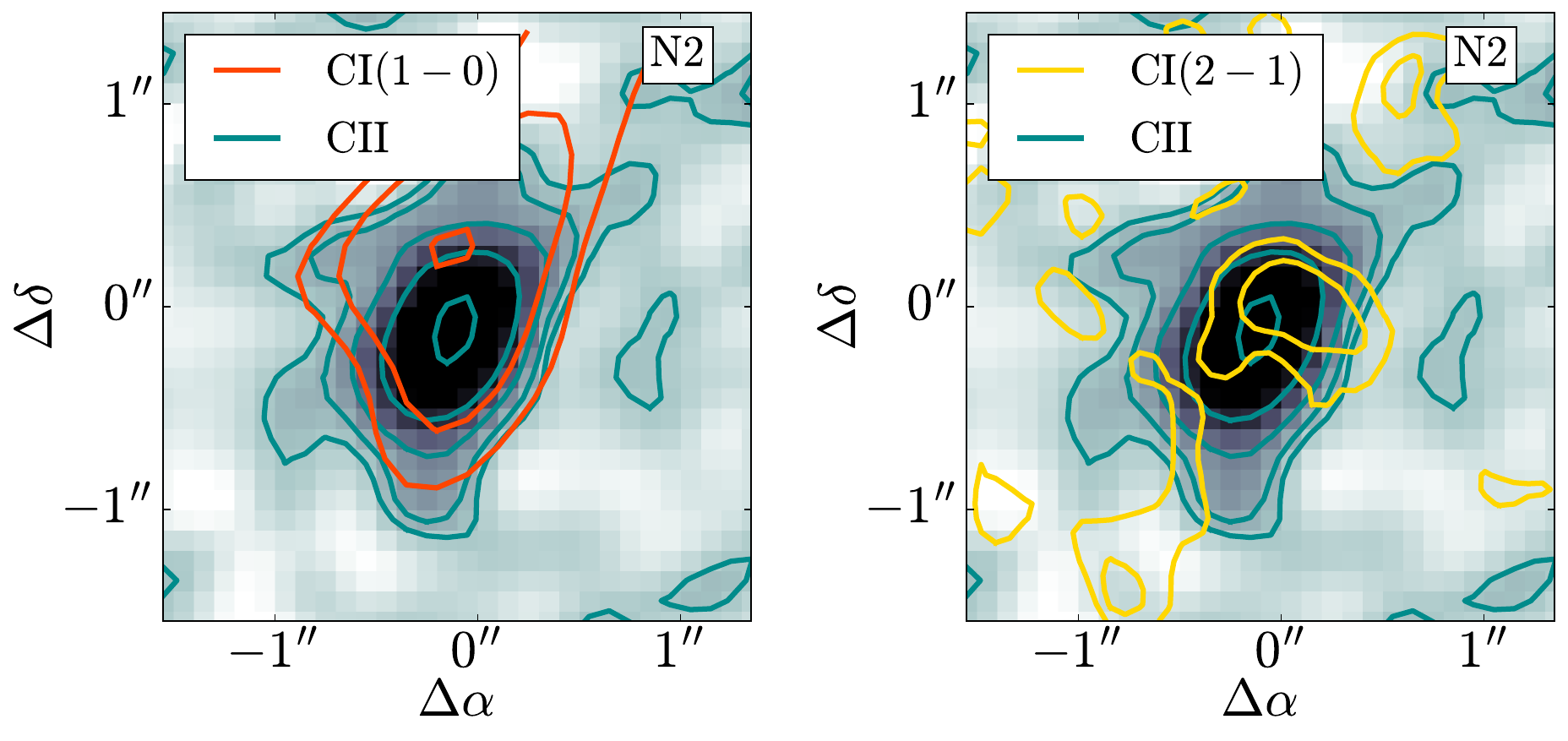}}}
    \caption{Continued.} 
\end{figure*}

\end{document}